\documentclass[prx,showpacs,twocolumn, floatfix,superscriptaddress,aps,10pt]{revtex4-2}

\usepackage[utf8]{inputenc}
\usepackage{graphics}
\usepackage{color}
\usepackage{stfloats}
\usepackage{amssymb}
\usepackage{amsmath}
\usepackage{overpic}
\usepackage{booktabs, array, mathptmx, tabularx, booktabs, lipsum, amsmath,multirow}
\usepackage{dcolumn}
\usepackage{epstopdf}
\usepackage{booktabs}
\usepackage{bm}
\usepackage{ulem}
\usepackage{graphicx}
\usepackage{subfigure}
\usepackage{threeparttable}
\usepackage{multirow}
\usepackage{xcolor}
\usepackage{makerobust}
\usepackage[caption=false]{subfig}

\newcommand{\captiontitle}[1]{\textbf{#1}}

\usepackage{xr}

\newcommand{\bvec}[1]{\boldsymbol{#1}}

\newcommand{\cidx}{\gamma}

\newcommand{\makeauthor}[2]{\newcommand{#1}[1]{{%
  \protect%
  \color{#2}{%
    \bfseries
  } ##1}}%
  \MakeRobustCommand#1}
\makeauthor{\af}{black}
\makeauthor{\qlxu}{black}
\usepackage{float}

\begin{document}

\preprint{APS/123-QED}

\title{Engineering 2D square lattice Hubbard models in 90$^\circ$ twisted Ge/SnX (X=S, Se) moir\'e superlattices }

\author{Qiaoling Xu}
\altaffiliation{These authors contributed equally.}
\affiliation{College of Physics and Electronic Engineering, Center for Computational Sciences, Sichuan Normal University, Chengdu 610068, China}
\affiliation{Tsientang Institute for Advanced Study, Zhejiang 310024, China}
\affiliation{Songshan Lake Materials Laboratory, 523808 Dongguan, Guangdong, China}

\author{ Ammon Fischer}
\altaffiliation{These authors contributed equally.}
\affiliation{Max Planck Institute for the Structure and Dynamics of Matter, Luruper Chaussee 149, 22761 Hamburg, Germany}

\author{Nicolas Tancogne-Dejean}
\affiliation{Max Planck Institute for the Structure and Dynamics of Matter, Luruper Chaussee 149, 22761 Hamburg, Germany}

\author{Tao Zhang}
\affiliation{Department of Materials Science and Engineering
City University of Hong Kong, Kowloon Tong,Hong Kong, SAR 999077, P. R. China}
\affiliation{Songshan Lake Materials Laboratory, 523808 Dongguan, Guangdong, China}

\author{Emil Vi\~nas Bostr\"om}
\affiliation{Max Planck Institute for the Structure and Dynamics of Matter, Luruper Chaussee 149, 22761 Hamburg, Germany}

\author{Martin Claassen}
\email[Email: ]{claassen@sas.upenn.edu}
\affiliation{Department of Physics and Astronomy, University of Pennsylvania, Philadelphia, PA 19104}

\author{Dante M. Kennes}
\email[Email: ]{Dante.Kennes@rwth-aachen.de}
\affiliation{Institut f\"ur Theorie der Statistischen Physik, RWTH Aachen University and JARA-Fundamentals of Future Information Technology, 52056 Aachen, Germany}
\affiliation{Max Planck Institute for the Structure and Dynamics of Matter, Luruper Chaussee 149, 22761 Hamburg, Germany}

\author{Angel Rubio}
\email[Email: ]{Angel.Rubio@mpsd.mpg.de}
\affiliation{Max Planck Institute for the Structure and Dynamics of Matter, Luruper Chaussee 149, 22761 Hamburg, Germany}
\affiliation{Initiative for Computational Catalysis and Center for Computational Quantum Physics, Simons Foundation Flatiron Institute, New York, NY 10010 USA}

\author{Lede Xian}
\email[Email: ]{xianlede@sslab.org.cn}
\affiliation{Tsientang Institute for Advanced Study, Zhejiang 310024, China}
\affiliation{Songshan Lake Materials Laboratory, 523808 Dongguan, Guangdong, China}
\affiliation{Max Planck Institute for the Structure and Dynamics of Matter, Luruper Chaussee 149, 22761 Hamburg, Germany}

\date{\today}

\begin{abstract}
Due to the large-period superlattices emerging in moir\'e two-dimensional (2D) materials, electronic states in such systems exhibit low energy flat bands that can be used to simulate strongly correlated physics in a highly tunable setup. While many investigations have thus far focused on moir\'e flat bands and emergent correlated electron physics in triangular, honeycomb and quasi-one-dimensional lattices, tunable moir\'e realizations of square lattices subject to strong correlations remain elusive. Here we propose a feasible scheme to construct moir\'e square lattice systems by twisting two or more layers of 2D materials in a rectangular lattice by 90 degrees. We demonstrate the concept with twisted GeX/SnX (X=S, Se) moir\'e superlattices and calculate their electronic structures from first principles. We show that the lowest conduction flat band in these systems can be described by a square lattice Hubbard model with parameters which can be controlled by varying the choice of host materials, number of layers, and external electric fields. In particular, twisted double bilayer GeSe realizes a square lattice Hubbard model with strong frustration due to the next nearest neighbour hopping that could host unconventional superconductivity{, in close analogy to the Hubbard model for copper-oxygen planes of cuprate high-temperature superconductors}. The presented scheme uses 90-degree twisted 2D materials with rectangular unit cells as a promising platform for realizing  the physical phenomena of square lattice Hubbard models, establishing a new route for studying its rich phase diagram of magnetism, charge order, and unconventional superconductivity in a highly tunable setting.

\end{abstract}

\maketitle

\section{\textbf{INTRODUCTION}}

The electronic Hubbard model \cite{Hubbard1,Kanamori1,Gutzweiler1} is ubiquitous in condensed matter research. In its original form it contains only two parameters: the electron hopping amplitude between adjacent sites (t) and a strong on-site interaction (U) between electrons of opposite spin. Despite its simplicity, the Hubbard model was employed early on to describe interaction-driven Mott metal-insulator transitions, with rich possibilities for magnetic order or quantum spin liquids at low temperatures \cite{RevModPhys.63.1,doi:10.1126/science.235.4793.1196}. Upon doping, the single band electronic Hubbard model is purported to support high-temperature superconductivity \cite{PhysRevLett.58.2794} and charge density waves \cite{doi:10.1126/science.aam7127,huang2018}, while strange metallicity and the pseudogap regime at finite temperatures have attracted considerable attention \cite{doi:10.1126/science.abh4273,Timusk_1999,doi:10.1146/annurev-conmatphys-090921-033948}.

From a theoretical perspective the apparent simplicity, yet intrinsic complexity, of the Hubbard model has captured the imagination of generations \cite{RevModPhys.66.763,RevModPhys.78.17,doi:10.1146/annurev-conmatphys-031620-102024}. For many state-of-the-art methods of quantum many-body physics the (approximate) solution of the Hubbard model is viewed as an essential benchmark. Many complementary theoretical advances \cite{doi:10.1126/science.aam7127,doi:10.1146/annurev-conmatphys-090921-033948,PhysRevX.5.041041}, such as dynamical mean-field theory, quantum Monte-Carlo, renormalization group based approaches or tensor network ans\"atze try to draw a picture as complete as possible for the different regimes of their validity. For example, one-dimensional Hubbard models are often amendable to tensor networks \cite{DMRG92,DMGR93}, while high-dimensional Hubbard models can be well described by a dynamical mean field approach \cite{RevModPhys.68.13}. The most pertinent case of the two-dimensional fermionic Hubbard model can be addressed on bipartite lattices and at half filling using a quantum Monte Carlo description \cite{PhysRevB.31.4403,PhysRevB.40.506,PhysRevB.94.085103,RevModPhys.77.1027}, but the superconducting state emerging at finite doping and frustration is difficult to capture with (numerically) exact methods beyond quasi-one-dimensional ladder geometries \cite{doi:10.1126/science.aal5304}. 
 
An alternative approach which has gained tremendous attention recently is to turn the problem upside-down from computation to (quantum) simulation: Instead of a theoretical approach or classical computation, cold gases or other platforms \cite{kennes_moire_2021} can provide highly controllable experimental simulations of the model system \cite{jordens2008mott,hart2015observation,mazurenko2017cold,gross2017quantum,singha2011two,hensgens2017quantum,PhysRevA.92.062318,stanisic2022observing}. Here, the simulation of the (fermionic or bosonic) Hubbard model again constitutes one of the main goals and benchmarks. Much progress has been achieved concerning quantum magnetism in Hubbard-type models utilizing such a simulation approach, but yet many questions -- e.g. the one of unconventional superconductivity -- remain beyond the current simulation capabilities due to the ultralow temperatures that are necessary to access the corresponding energy scales \cite{xu2023frustration}.   

With the recent rise of ultra-clean and highly controllable two-dimensional materials, a new protagonist has entered the stage \cite{novoselov20162d,lemme20222d,ares2022recent,choi2022large,castellanos2022van}. These materials are unique as they provide a high degree of tunability via heterostructure stacking, gating, and stress or strain, in a low disorder setting. Using stacking configurations that include either a relative twist between adjacent layers or an intrinsic lattice constant mismatch between layers of different materials, such moir\'e heterostructures have recently become one of the main research focus of the field \cite{kennes_moire_2021,cao2018unconventional,cao2018correlated,choi2019electronic,kerelsky2019maximized,andrei2020graphene,wang2020correlated,xu2020correlated,mak2022semiconductor,zeng2023thermodynamic,park2023observation,xu2023observation}. The large scale moir\'e interference pattern emerging at low twist angle or small lattice constant mismatches allows to quench the kinetic energy scales promoting competing energy scales (such as the potential, spin-orbit or phononic couplings). This concept has been explored extensively mainly in graphitic or transition metal chalcogenide materials and has raised the hope to realize many of the prototypical models of condensed matter physics \cite{bistritzer2011moire,wu2018hubbard,tang2020simulation,xian2021realization,kennes_moire_2021,xu2022tunable,kennes2020one,wang2022one,claassen2022ultra,xian2021engineering}. In this work, we propose moir\'e materials engineering as an inroad to realize the prototypical square lattice Hubbard model with a nearest neighbor hopping $t$ and next nearest neighbor hopping $t'$. In contrast to the usual paradigm of using small twist angles we study a bilayer system with each layer exhibiting a rectangular but only slightly non-square unit cell rotated by 90$^\circ$ with respect to each other. The small lattice constant mismatch of the two sides of the rectangular unit cell combined with the rotation naturally generates a moir\'e pattern giving rise to the square lattice structure.  
This provides a pivotal materials-based platform to study unconventional superconductivity in a highly controlled two-dimensional materials platform allowing to bridge the gap  left open between theoretical approaches and quantum simulations in the future.

\section{\textbf{Moir\'e square superlattices from twisted rectangular lattices}}

\begin{figure} [h]
\includegraphics[width=3.5in]{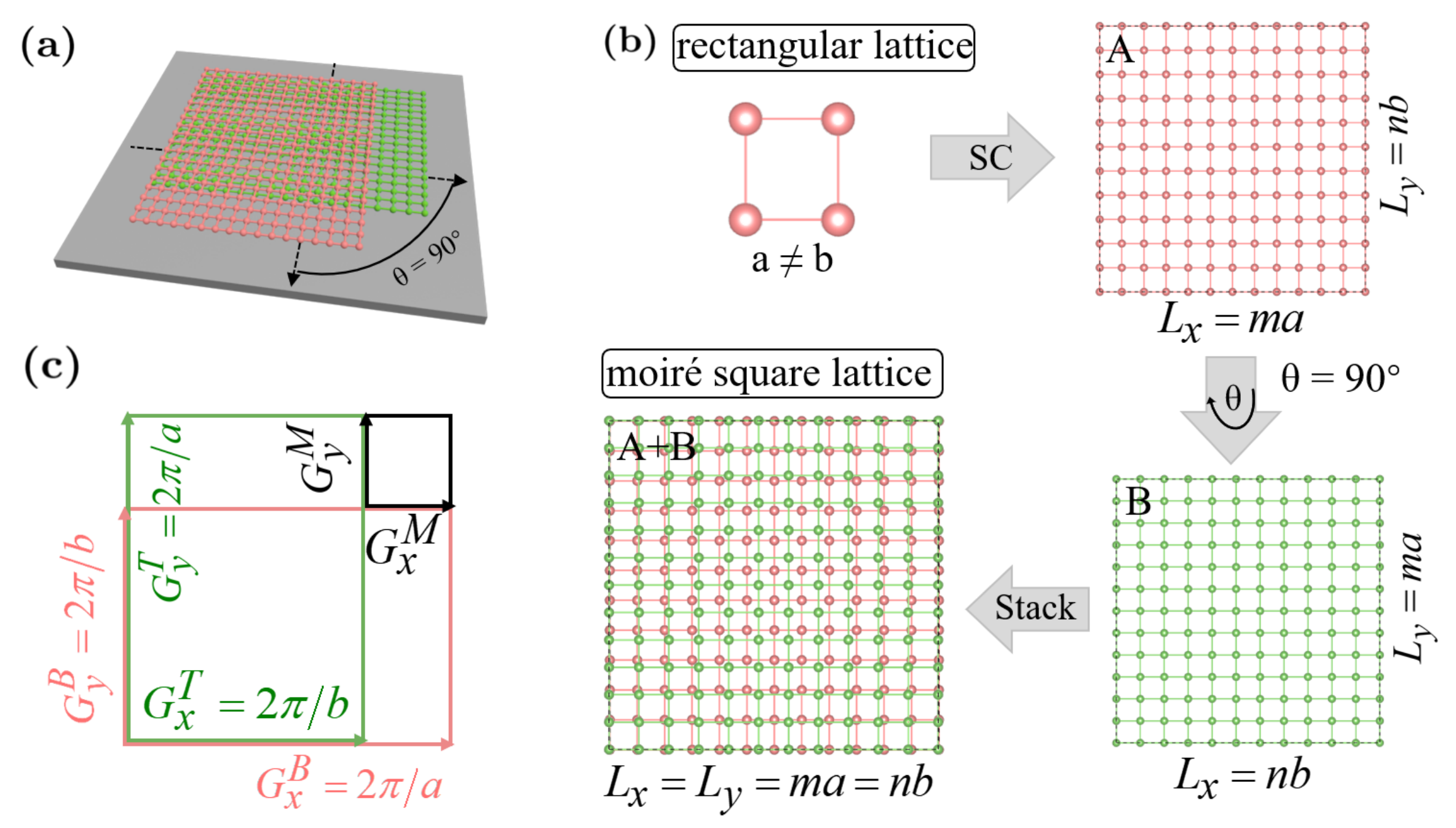}
\centering
\caption{ \captiontitle{General strategy of constructing moir\'e square lattices using 2D rectangular lattices. } (a) Schematic illustration of the 90$^\circ$ twisted 2D layers. (b) Flow chart of the construction of the moir\'e square superlattice using 2D layers of rectangular lattices with different lattice constants a and b (a $\neq$ b). (c) Brillouin zone of the 90$^\circ$ twisted rectangular lattices. }
\label{Fig1}
\end{figure}

\subsection{\textbf{General strategy}}

The natural way to obtain a 2D moir\'e square lattice is through twisting two layers of 2D materials with a square lattice unit cell. However, in practice, the number of 2D layered materials with square lattices that can be synthesized or exfoliated down to monolayer or few layers in experiments is very limited. On the other hand, thin layers of 2D materials with rectangular lattices are more commonly studied, such as the group-IV monochalcogenides GeX/SnX (X=S, Se, Te) family \cite{chang2020experimental}, $\alpha$-antimonene and $\beta$-tellurene \cite{wang2018field,shi2019van,shi2020tuning}, as detailed in Table S1 of the supplementary materials (SM). Herein, we propose a general strategy to construct moir\'e square lattices basing on the orthorhombic van der Waals (vdW) crystal structures with rectangular unit cell, as shown in Figure 1. 

Starting with two identical layers of 2D materials in a simple rectangular lattice with lattice constants $a\neq b$, one can twist one of the layers by $90^\circ$ and stack it on top of the other layer. In this way, the lattice constants of the top and the bottom layers along the same axis are no longer equal and a moir\'e pattern will emerge. In the real space, the moir\'e wavelength $L$ along the x and the y axis are both determined by the lattice mismatch between the original lattice constants $a$ and $b$, i.e., $L_x=L_y=ma=nb$ (see Figure~\ref{Fig1}). Therefore, the moir\'e period along the x and the y axis are equal and the resultant moir\'e pattern becomes a square lattice. In reciprocal space, as shown in Figure~\ref{Fig1}(c), the reciprocal lattice vectors of the top/bottom layer and those of the moir\'e superlattice are connected by the following relation:
\begin{equation}
    G_x^B-G_x^T = G_x^M,
\end{equation}
where $G_x^B=2\pi/a$, $G_x^T=2\pi/b$ and $G_x^M=2\pi/L$ are the reciprocal lattice vectors of the bottom layer, the top layer and the moir\'e superlattice, respectively. Thus, real space lattice vectors $a$, $b$ and $L$ satisfy the following equation:
\begin{equation}
    \frac{1}{a}-\frac{1}{b} = \frac{1}{L}.
\end{equation}
So $L$ can be determined by the expression $L=[(1/a)-(1/b)]^{-1}$. The smaller the difference between the original lattices constants $a$ and $b$, the larger the moir\'e period $L$. In practice, we will need to slightly strain the original lattice constants $a$ and $b$ such that we can construct a commensurate supercell containing only one moir\'e wavelength (see the Table~\ref{tab:table1}). 
This method of straining lattice constants to achieve commensurability was also very successful in the past to construct moir\'e triangular lattices using lattice mismatches of TMD heterostructures \cite{tang2020simulation,regan2020mott,xu2020correlated,lau2022reproducibility}, as further demonstrated in Fig. S2.

The fundamental design principles for constructing a square moir\'e lattice with isolated flat bands using rectangular unit cells are twofold. First, the chosen material should be semiconducting or semimetallic. This choice facilitates experimental control over the Fermi level and carrier concentration through gating. Moreover, the moir\'e potentials originated from the interlayer coupling are relatively weak, which are the most effective to the electronic states at the band edges. If the system is metallic, not all the electronic states around the Fermi level can be modified by the moir\'e potentials and those that are modified may be mixed with bulk metallic states, preventing the formation of clean, isolated flat bands needed for quantum simulations.  Second, the material should have an appropriate lattice mismatch, with the resulting supercell lattice constant in the range of 10 $\sim$ 100 \AA. If the lattice mismatch is too large, leading to a small supercell period, the moir\'e potentials cannot effectively confine low-energy electronic states. For instance, materials like 1T'-WTe$_2$, despite having a rectangular unit cell, do not have sufficient lattice mismatch between the two lattice vectors, making it difficult to achieve isolated flat bands in a 90$^\circ$ twisted configuration.  Conversely, if the lattice mismatch is too small, resulting in an overly large supercell lattice constant, maintaining structural homogeneity in experiments becomes challenging, which can disrupt flat band formation in the system (see Table S1 in the SM). 

Compared to other possible approaches for generating superlattices with patterned substrates, our proposed system features moir\'e wavelengths below 10 nm, which are significantly smaller than the superlattice wavelengths reported in previous studies. For example, wavelengths close to 40 nm have been reported in the literature \cite{forsythe2018band}. The reduced moir\'e wavelength leads to substantially stronger Hubbard interactions, which are inversely proportional to the wavelength. The stronger Hubbard interaction \cite{Hubbard1} facilitates the observation of correlation effects at experimentally accessible temperatures, broadening the range for tunability. Additionally, it is much easier to maintain structural homogeneity with a smaller supercell wavelength, which is also beneficial for experimental observations.

\subsection{\textbf{90$^\circ$ twisted bilayer GeX/SnX (X=S, Se)}}

\begin{figure}[h]
\includegraphics [width=3.3in] {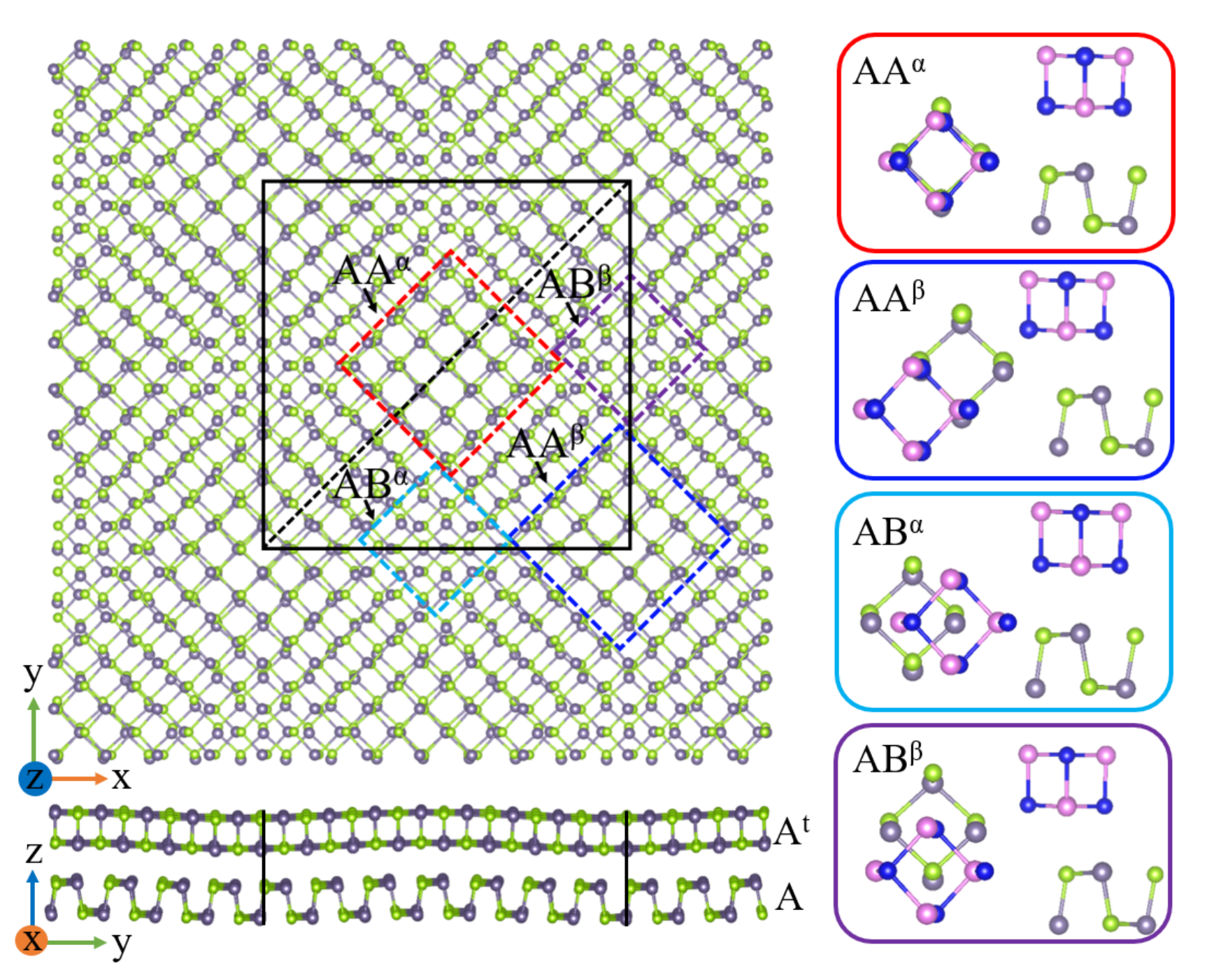}
\centering
\caption{\captiontitle{Moir\'e structure of 90$^\circ$ twisted bilayer GeSe.} The top and the side views of the atomic structure are shown in upper left and lower left panels, respectively. A$^t$ in the side view is used to indicate the layer twisted by 90$^\circ$ with respect to the layer labeled as A. The moir\'e unit cell is indicated by black solid lines. The black dashed line represents the in-plane $C_{2d}$ rotational axis. The top and the side views of the atomic structures of the local stacking configurations of AA$^\alpha$, AA$^\beta$, AB$^\alpha$, and AB$^\beta$ are shown in right panels. The Ge and the Se atoms are represented by green/blue and grey/pink balls, respectively.}
\label{Fig2}
\end{figure}

We first apply this strategy to construct moir\'e square superlattices with 2D GeSe that have a rectangular lattice structure similar to that of black phosphorene \cite{wiedemeier1978refinement}. 
The optimized atomic structures of 90$^\circ$ twisted bilayer GeSe is shown in Figure~\ref{Fig2}. As expected, the twisted structure exhibits a moir\'e square superlattice with moir\'e lattice constant of 25.658 \AA. The detailed structural parameters are summarized in Table~\ref{tab:table1}. The moir\'e structure mainly consists of two types of AA-like and two types of AB-like local stacking registries called AA$^\alpha$, AA$^\beta$, AB$^\alpha$ and AB$^\beta$. Each of these is uniquely identified by which atoms fall on top of each other in a top and side view as categorized in the right panels of Figure~\ref{Fig2}. 
These four special stackings can be transformed into each other by translational sliding of one of the basal planes in the unit cell. 
Moreover, for such twisted bilayer structures, not only are the lattice constants along the two orthogonal directions equal, but the system also has an in-plane $C_{2d}$ rotational symmetry along the diagonal direction of the moir\'e cell (see Figure~\ref{Fig2} dotted line), leading to equivalent electronic properties along the x and the y directions.  

\begin{figure}[h]
\includegraphics [width=3.5in] {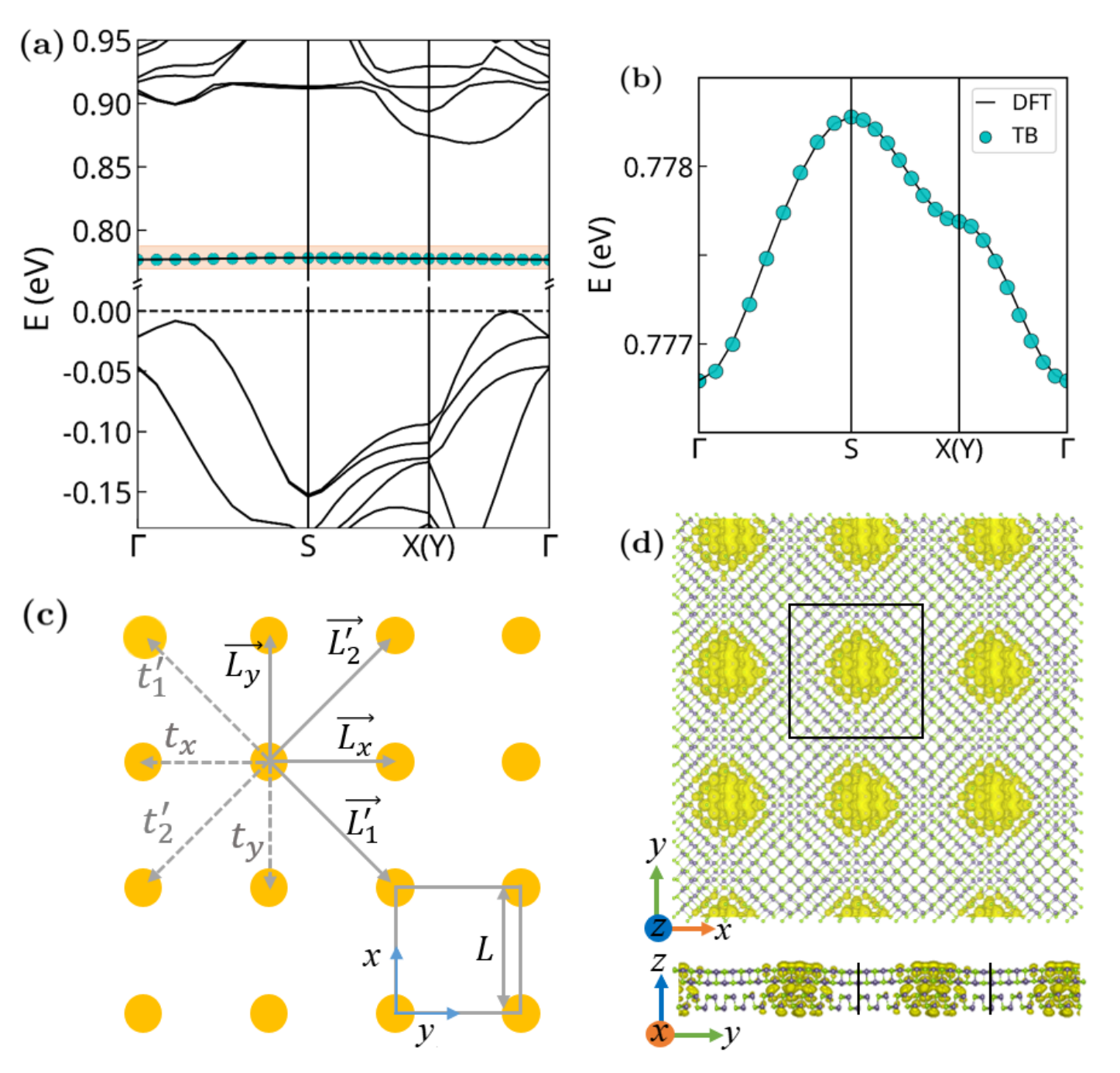}
\centering
\caption{\captiontitle{Moir\'e flat bands of 90$^\circ$ twisted bilayer GeSe.} (a) Low-energy band structure near the band edge. The dashed line indicates the Fermi level. (b) The enlarged band structure corresponding to the shaded region in (a), which shows the dispersion of the moir\'e flat band at the CBE. The black solid line and the cyan dots denote the results calculated with DFT and the effective TB model, respectively. (c) Schematic diagram of the square lattice effective TB model. (d) Top view (upper panel) and side view (lower panel) of partial charge density distribution in real space for the flat-band state at the CBE.}
\label{Fig3}
\end{figure}

\begin{table}[h]
    \centering
    \caption{\captiontitle{Parameters of constructing square lattices for 90$^\circ$ twisted bilayer GeX/SnX.} (a,b) denotes the two lattice constants in the primitive unit cell; (m, n) and (m$^{\prime}$, n$^{\prime}$) denote the supercell matrix for the top and the bottom layers, respectively; p and q denote the applied stress on lattice constants a and b, respectively, to achieve commensurability.}
    \label{tab:table1}
  \begin{tabular}{|c||c|c|c|c|}
    \hline\hline
    Materials & Unit cell (a, b)/{} \AA & (m, n) & (m$^{\prime}$, n$^{\prime}$)  & Scale (p, q)/$\%$ \\
    \hline\hline
    GeS &
        (3.64, 4.29) & (7, 6) & (6, -7) & (0.5, -0.5) \\
    GeSe &
        (3.83, 4.38) & (8, 7) & (7, -8) & (0.08, -0.08) \\
    SnS &
        (3.98, 4.33) & (12, 11) & (11, -12) & (-0.2, 0.2) \\
    \hline\hline
  \end{tabular}
\end{table}

\begin{figure}[h]
\centering
\includegraphics [width=3.0in] {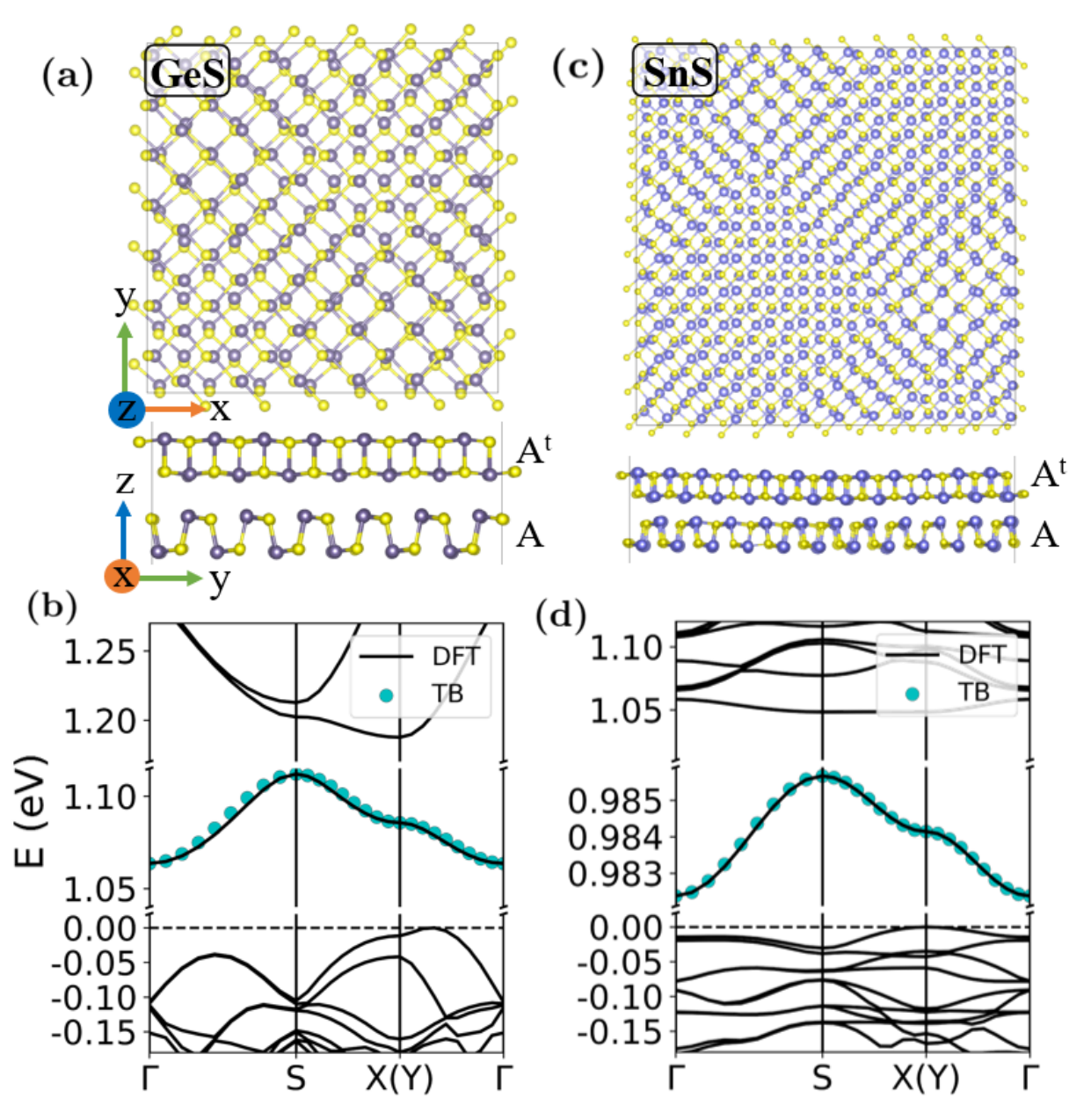}
\caption{ \captiontitle{Moir\'e flat bands of 90$^\circ$ twisted bilayer GeS and SnS.} The atomic and the electronic structures of 90$^\circ$ twisted bilayer GeS (a, b) and SnS (c, d).  The low energy flat bands at the CBE are fitted well by the effective TB model.}
\label{Fig4}
\end{figure}

We then calculate the electronic band structures for a 90$^\circ$ twisted bilayer GeSe system using density functional theory (DFT) calculations. The result is shown in Figure~\ref{Fig3}(a). As expected, the band structure of the system along the X and the Y directions are identical. Due to the moir\'e modulation, an isolated flat band with ultra-small bandwidth of 1.5 meV emerges at the conduction-band edges (CBE). The calculated charge density distribution of the flat band shows that these electronic states are localized at the AA$^\alpha$ regions, forming a square checkerboard pattern in real space (see Figure~\ref{Fig3}(d)).  Based on these results, we model the flat band with a effective tight-binding (TB) model on a square lattice for which the Hamiltonian can be written as: 
\begin{equation}
    \mathbf{H} = - \sum_{{\left \langle {i,j} \right \rangle},\sigma} t_{ij}c^\dagger_{i\sigma}c^{\phantom\dagger}_{j\sigma}  + 
  \sum_{\left \langle \left \langle {i,j} \right \rangle \right \rangle}
      t^{\prime}_{ij}c^\dagger_{i\sigma}c^{\phantom\dagger}_{j\sigma},
\end{equation}
where $t_{ij}$ and $t^{\prime}_{ij}$ are the nearest-neighbour and the next-nearest-neighbour hopping amplitudes, respectively; ${\left \langle...\right \rangle}$ and ${\left \langle \left \langle ...\right \rangle \right \rangle}$ represent the summation over the nearest-neighbours and next-nearest-neighbours, respectively; $c_{i\sigma}$ (c$^\dagger_{i\sigma}$) annihilates (creates) an electron at site $i$ with spin $\sigma = \uparrow$, $\downarrow$. As the spin-orbit coupling for the flat band is negligible and thus the spins can be regarded as degenerate, using the notations shown in Figure~\ref{Fig3}(c), the 
dispersion relation for one of the spin component can written as:
\begin{multline}
    E (k) = 2t_x\cos{k_{x}L} + 2t_y\cos{k_{y}L}  \\
    + 2t^{\prime}_1\cos{(k_{x}L - k_{y}L)} + 2t^{\prime}_2\cos{(k_{x}L + k_{y}L)},
\label{eq:tb}
\end{multline}
where $t_x$($t_y$) denotes the nearest-neighbour hopping amplitude along the $x(y)$ direction and $t^{\prime}_{1}/t^{\prime}_2$ the next-nearest-neighbour hoppings along the diagonal directions; $L$ denotes the lattice constant for the moir\'e square superlattice. As shown in Figure~\ref{Fig3}(b),  the fitted effective TB model gives almost identical band dispersion as the one calculated with DFT for the flat band at the CBE in the 90$^\circ$ twisted bilayer GeSe with $t_x = t_y$ (see Table~\ref{tab:table2} for values of the fitted parameters). 

We further apply the strategy to other 2D group-IV monochalcogenides GeS and SnS. Although the lattice constants of GeS and SnS are different from those of GeSe, the generality of the strategy ensures that the moir\'e structures end up as square superlattices. The atomic structures of the 90$^\circ$ twisted bilayer GeS and SnS are shown in Figures~\ref{Fig4}(a) and~\ref{Fig4}(c) and the detail parameters are listed in Table~\ref{tab:table1}. The moir\'e patterns of these systems are mainly formed by four local stacking domains similar to those of twisted bilayer GeSe. The corresponding DFT band structures are shown in Figures~\ref{Fig4}(b) and~\ref{Fig4}(d). Similar to twisted GeSe, a flat bands with narrow bandwiths of 48 meV and 3 meV appear at the CBE for twisted bilayer GeS and SnS, respectively. As shown in Figure~\ref{Fig4}, these flat bands can be accurately fitted by the effective square lattice model described in Equation \eqref{eq:tb} and we provide the hoppings amplitudes in Table~\ref{tab:table2}.       

The moir\'e potentials induced by the interlayer hybridization variation across the superlattice period are mainly responsible for the formation of flat bands and charge density localization behavior at the square lattice sites. To illustrate this effect, we further perform two DFT calculations: one including only interlayer hybridization without atomic relaxation, and another with atomic relaxation fixed to that of the fully relaxed moir\'e bilayer, but with one layer removed to eliminate interlayer hybridization. The results are presented in Fig. S1. As shown in the figure, even with interlayer hybridization alone, ultraflat moir\'e minibands appear at the band edges. However, when interlayer hybridization is artificially removed by omitting one layer—while maintaining atomic reconstruction—this setup is insufficient to sustain flat bands at the band edges.

\begin{figure}[h]
\centering
\includegraphics [width=0.8\columnwidth] {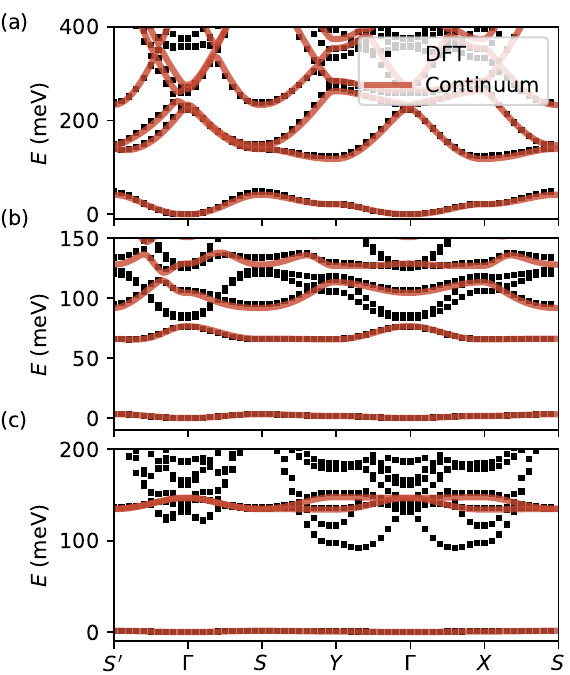}
\caption{ \af{\captiontitle{$\Gamma$-valley continuum model for 90$^\circ$ twisted bilayer GeX/SnX (X=S, Se).} Continuum model and \textit{ab-initio} bandstructure for the square lattice moir\'e candidates (a) GeS, (b) SnS and (c) GeSe.}}
\label{fig:continuum_vs_dft}
\end{figure}

\subsection{\af{\textbf{$\Gamma$-valley physics and building an effective continuum description}}}
\af{The flat moir\'e bands in 90$^\circ$ twisted bilayer GeX/SnX (X=S, Se) are shaped by states from the $\Gamma$-valley of the untwisted bilayers. 
Unfolding the full-fledged DFT bandstructure from the mini-Brillouin zone of the supercell to the primitive Brillouin zone of the pristine bilayers, see Fig.~S7 for the case of GeSe, reveals that the flat band of the CBM is primarily constituted by a linear combination of Bloch states near the $\Gamma$-point of the primitive Brillouin zone.}

\af{
The unfolding analysis of Bloch states allow us to construct simplified analytical continuum models that describe the effective $\Gamma$-valley physics in 90$^\circ$ twisted bilayer GeX/SnX (X=S, Se).
%
%
Within aligned bilayers featuring the same local AA/AB$^{\alpha, \beta}$ stacking that appears in the twisted configuration, the bands at the $\Gamma$-point are primarily shaped by the Ge p$_z$-orbitals, see Supplementary Material Fig. S8 for orbital-projected bandstructures of the respective bilayer configurations.
The out-of-plane orientation of these orbitals leads to strong inter-layer hybridization such that the energetically lowest states in the CBM are composed of bonding $p_z$ states, whereas anti-bonding states are well-separated in energy by $\sim 1.5$ eV.
%
%
Within our low-energy theory, we therefore only retain the bonding state at $\Gamma$, described by a $\bvec k\cdot\bvec p$ Hamiltonian~\cite{angeli2021gamma,fujimoto2021effective} of the form
\begin{equation}
    H(\bvec k) = \frac{\bvec k^2 }{ 2m^* } + \bvec \Delta(\bvec r) \,.
    \label{eq:gamma-continuum}
\end{equation} 
where $m^*$ is the effective mass associated to the bonding $p_z$-state of the GeX/SnX (X=S, Se) bilayers and $\Delta(\bvec r)$ is the moir\'e potential felt by the electrons in the CBM as a function of the real-space position $\bvec r$ in the square moir\'e unit cell.
The moir\'e potential is approximated in lowest harmonic Fourier expansion
\begin{align}
    \Delta(\bvec r) &{}=  \sum_s \sum_{j=1}^4 V(\bvec g^s_j) \exp(i [\bvec g^s_j \cdot \bvec r + \psi^s_j]) + \text{h.c.} \,,
    \label{eq:moire-potential}
\end{align}
where $\bvec g^s_j$ are the four reciprocal lattice vector in shell $s$ of the moiré Brillouin zone sorted by increasing distance $|\bvec g^s_j|$.
The in-plane rotational symmetry $C_{2d}$ enforces $\Delta(\bvec r) = \Delta(C_{2d} \bvec r)$, which restricts possible values of $(V(\bvec g^s_j), \psi^s_j)$. Their magnitude is extracted by fitting to the \textit{ab-initio} DFT bandstructure and we give a detailed overview of symmetry-allowed continuum model parameters in the Supplementary Material.}

\af{
Our results are shown in Fig.~\ref{fig:continuum_vs_dft}~(a)-(c).
The continuum model is able to reproduce not only the flat moir\'e band, but also the higher energy bands in the conduction band manifold to good accuracy. For twisted bilayers of SnS and GeSe, additional bands admix to the higher-lying conduction bands, which do not originate from bonding states at $\Gamma$. However, these states do not constitute the moir\'e flat band.
}

\begin{figure}[h]
\centering
\includegraphics [width=3.4in] {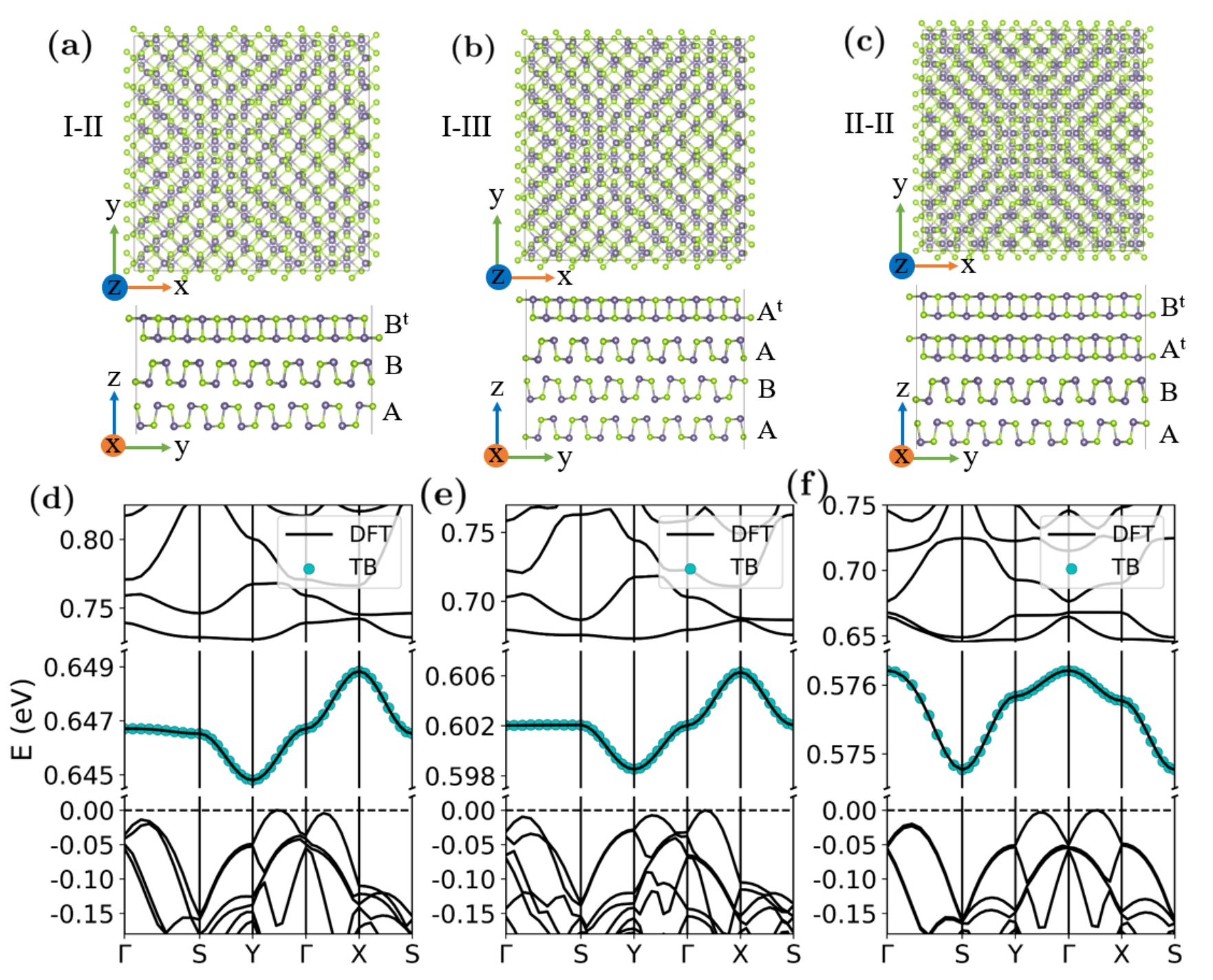}
\caption{\captiontitle{Moir\'e flat bands of 90$^\circ$ twisted multilayer GeSe.} (a-c) Top (upper panel) and side views (lower panel) of the atomic structures for the 90$^\circ$ twisted monolayer-bilayer (I-II), monolayer-trilayer (I-III) and AB-A$^t$B$^t$ type of bilayer-bilayer (II-II) configurations, where AB,  A$^t$ and B$^t$ indicate natural bilayer from bulk, $90^\circ$ twisted A layer and $90^\circ$ twisted B layer, respectively. The grey and green balls represent Ge and Se atoms, respectively. (d-f) The corresponding low energy band structures with fitting by the effective TB model.} 
\label{Fig5}
\end{figure}

\subsection{\textbf{90$^\circ$ twisted multilayer GeX/SnX (X=S, Se)}}

The general strategy we proposed can also be applied to twisted multilayer GeX/SnX (X=S, Se). With the additional layer degree of freedom, we can further engineer different flat band dispersion in a moir\'e square lattice. According to the review by Kai Chang and Stuart S.P. Parkin \cite{chang2020experimental} and the 2D materials database from high-throughput computational exfoliation of experimentally known compounds \cite{mounet2018two}, the exfoliation energy of GeS, GeSe and SnS are around $28\sim 36$ eV/{\text{\AA}}$^2$, comparable to that of black phosphorene ($30\sim 38$ eV/{\text{\AA}}$^2$). Black phosphorene has been successfully exfoliated down to monolayer \cite{li2017direct} and twisted monolayer-bilayer phosphorene has been fabricated in experiments \cite{zhao2021anisotropic}. With advancements in exfoliation technique for 2D layers developed in recent years \cite{guo2023ultra, wang2023clean}, we expect that all three systems (GeS, GeSe and SnS) considered in this work can be further exfoliated down to monolayer in experiments. Currently, thin films of these materials with 2 to 4 layers have already been successfully obtained using liquid exfoliation methods. Taking the GeSe structures as typical examples, the multilayer moir\'e square superlattices can be generated by twisting monolayer on pristine bilayer (I-II), on pristine trilayer (I-III) or bilayer on bilayer (II-II) as shown in Figures~\ref{Fig5}(a-c). The untwisted layers remain in their natural AB stacking in their pristine bulk form. The corresponding band structures for these moir\'e superlattices are given in Figures~\ref{Fig5}(d-f). Similar to the twisted bilayer systems, all of these twisted multilayer structures exhibit very flat minibands appearing at the CBE with bandwidths of 4 meV, 7 meV and 1.4 meV, respectively. In these three types of twisted multilayer structures, the in-plane $C_{2d}$ rotational symmetry is broken and the electronic properties along the x and the y axis are not equivalent. Nevertheless, the bottom moir\'e flat bands at the CBE in the twisted multilayer superlattices can also be described by the simple square lattice effective model given by Equation~\eqref{eq:tb} with different values for t$_x$ and t$_y$ (see fitted parameters in Table~\ref{tab:table2}). Therefore, with ultraflat bands and inequivalent hoppings along the x and the y axis, these moir\'e systems can be used to realize anisotropic square lattice Hubbard models.  

\begin{figure}[h]
\centering
\includegraphics [width=3.0in] {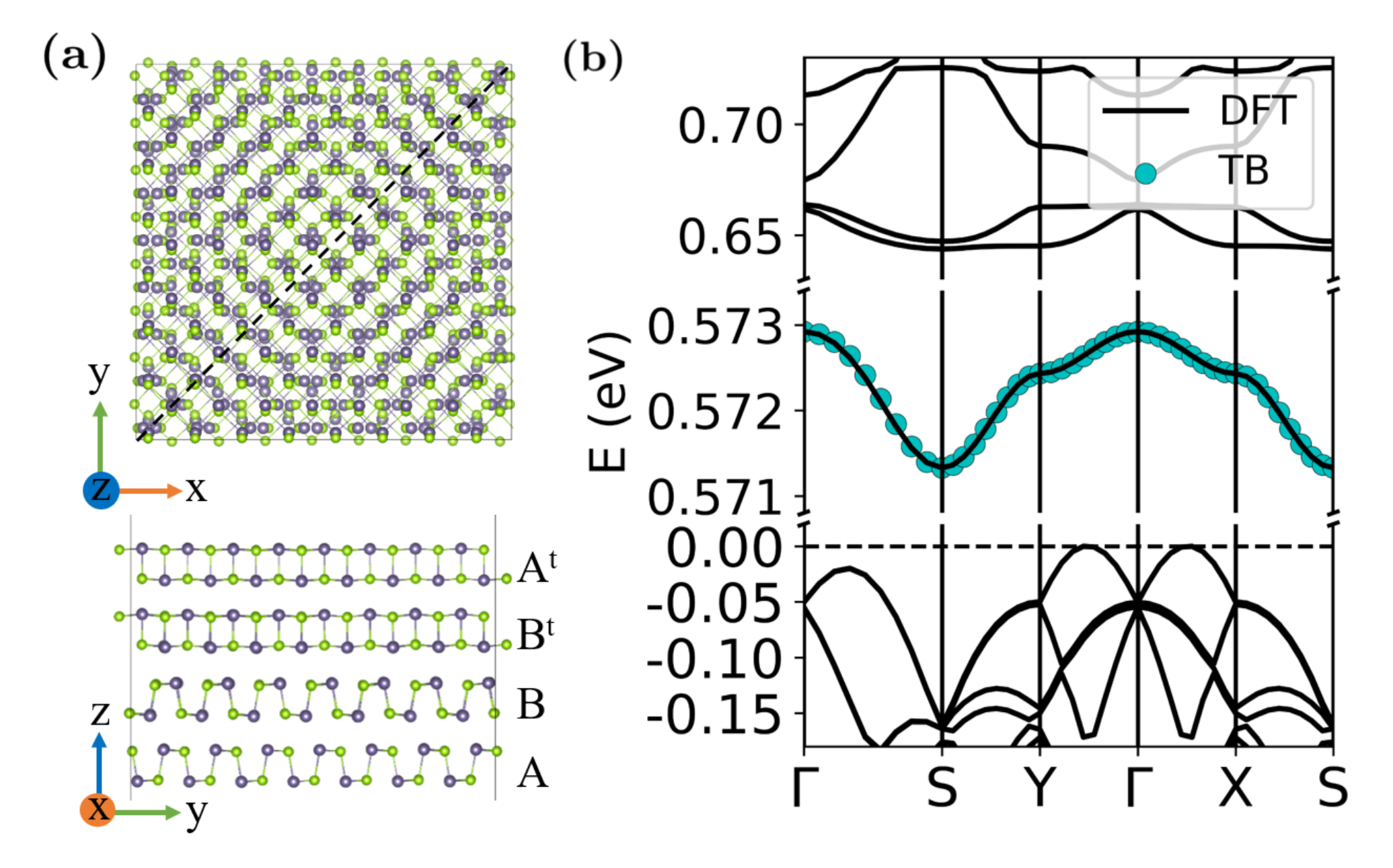}
\caption{ \captiontitle{90$^\circ$ twisted double bilayer GeSe with symmetry.} The atomic (a) and electronic structure (b) of the AB-B$^t$A$^t$ type of twisted double bilayer GeSe where the $C_{2d}$ symmetry around the diagonal axis of the moir\'e supercell is preserved. The low-energy flat band at CBE is fitted by the effective TB model shown in green markers.}
\label{Fig6}
\end{figure}

For twisted multilayer systems with even number of layers, the in-plane $C_{2d}$ rotational symmetry can be restored by designing the twist stacking. For example, in the twisted double bilayer system, by twisting the top two layers by 90$^\circ$ with respect to the bottom two layers, one can obtain a moir\'e square superlattice in the AB-A$^t$B$^t$ type of stacking (Figure~\ref{Fig5}(c)). By further flipping the top two layers up side down in the AB-A$^t$B$^t$ type of twisted double bilayer superlattice, one obtains the AB-B$^t$A$^t$ type of moir\'e superlattice (Figure~\ref{Fig6}(a)). While the in-plane $C_{2d}$ rotational symmetry along the diagonal direction of the supercell is absent in the AB-A$^t$B$^t$ type of moir\'e superlattice, it is restored in the the AB-B$^t$A$^t$ type. As shown in Figure~\ref{Fig6}(b), the band dispersion for the AB-B$^t$A$^t$ type twisted double bilayer GeSe along the x and the y directions are exactly the same due the symmetry in the system. In particular, the lowest moir\'e flat band at the CBE can be fitted to the effective TB model with $t_x=t_y$ (see Table~\ref{tab:table2}). Interestingly, for twisted multilayer moir\'e square superlattices, the next nearest neighbour hopping versus nearest neighbour hopping ratio $t^{\prime}/t$ is generally larger than that of the twisted bilayer. For the AB-B$^t$A$^t$ type twisted double bilayer GeSe, this value reaches $\sim 20\% $.

\subsection{\textbf{near 90$^\circ$ twisted GeX/SnX (X=S, Se)}}
\begin{figure}[h]
\centering
\includegraphics [width=3.5in] {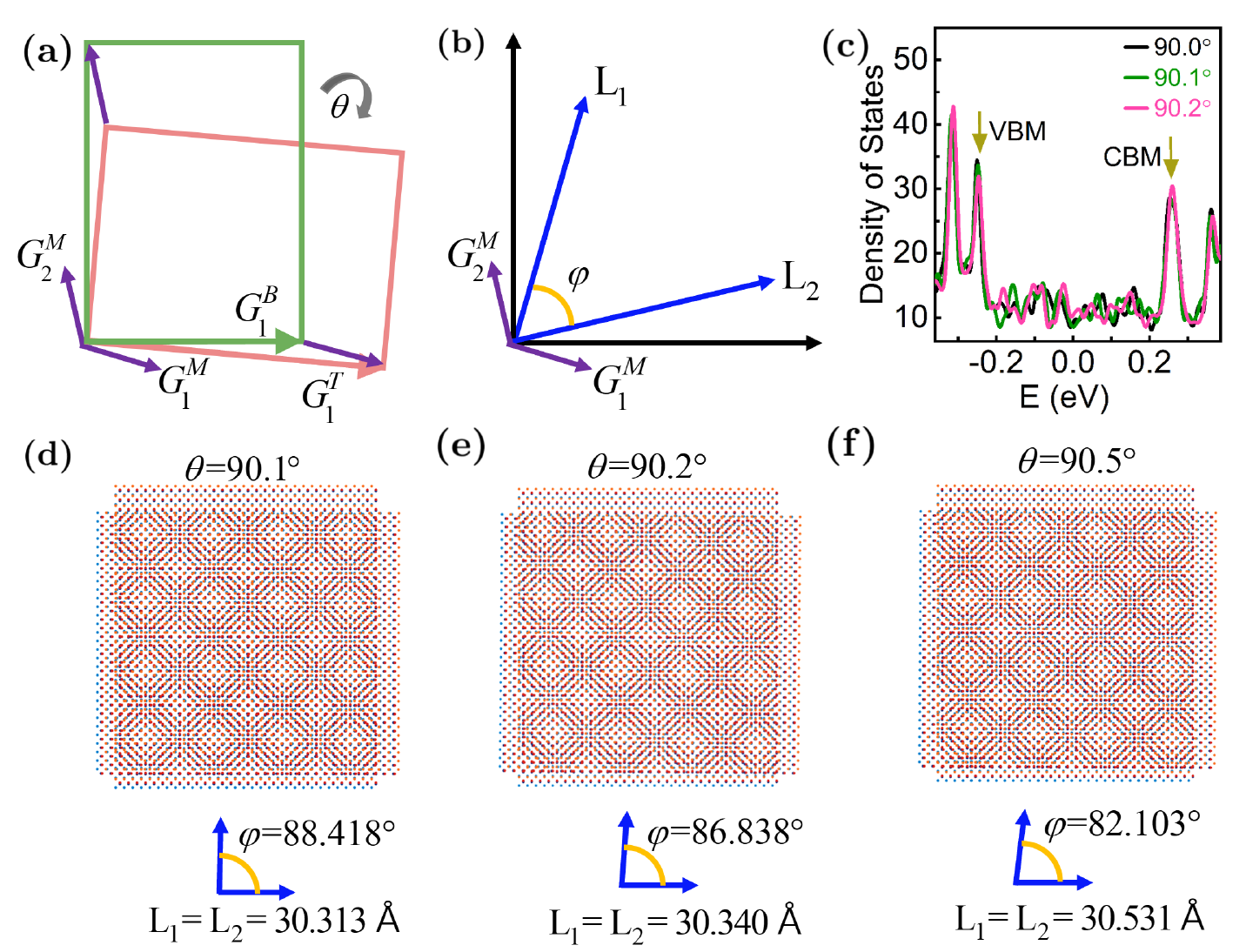}
\caption{\captiontitle{Nearly 90$^\circ$ twisted bilayer GeSe.} (a) Brillouin zone of nearly 90$^\circ$ twisted rectangular lattices with the reciprocal lattice vectors G$_1^M$ and G$_2^M$. (b) Relationship between the reciprocal lattice vectors and real space supercell lattice vectors. (c) Density of states of twisted bilayer GeSe at different twist angles near 90$^\circ$ using DeepH. (d-f) The atomic structures of 90.1, 90.3, 90.5$^\circ$ twisted GeSe in real space with their estimated supercell lattice vectors L$_1$ and L$_2$ and the angle between them shown in the lower panels.}
\label{Fig7}
\end{figure}

When one of the layers is twisted by 90$^\circ$ deviated by a small angle $\Delta\theta$ and subsequently stacked, a rhombus-shaped moir\'e pattern forms, similar to those that observed in the system with a perfect 90$^\circ$ twist. In reciprocal space, as shown in Figure~\ref{Fig7}(c), the relationship between the reciprocal lattice vectors of the top and bottom layers and those of the moir\'e superlattice is given by:
\begin{equation}
\bm{G}_{i}^B - \bm{G}_{i}^T = \bm{G}_{i}^M, \quad (i = 1, 2) 
\end{equation}
where \(\bm{G}_{1,2}^B\), \(\bm{G}_{1,2}^T\), and \(\bm{G}_{1,2}^M\) are the reciprocal lattice vectors of the bottom layer, top layer, and moir\'e superlattice, respectively. Thus, the real space lattice vector \(\bm{L}_{1,2}\) can be derived from the relationship \(\bm{G}_{i}^M \cdot \bm{L}_{i} = 2\pi \; (i = 1, 2)\). The angle $\theta$ between \(\bm{G}_1^M\) and \(\bm{G}_2^M\), the angle $\varphi$ between \(\bm{L}_1\) and \(\bm{L}_2\), and the magnitudes of each vectors for near 90$^\circ$ twisted GeSe are listed in SM Table S2.
Using the small angle approximation, one can show that when the twist angle $\theta$ is deviated from 90$^{\circ}$ by $\Delta\theta$, the angle $\varphi$ between the superlattice vectors will deviate from 90$^{\circ}$ by $\Delta\varphi=(\frac{a+b}{b-a})\Delta\theta$, where a, b are the two primitive cell lattice constants. Therefore, a slight deviation of the twist angle from 90$^\circ$ results in a magnified deviation of the angle between the supercell lattice vectors from 90$^\circ$. The magnified ratio is proportional to the sum of the primitive cell lattice constants, $a+b$, divided by the lattice mismatch, $b-a$. Nevertheless, for 2D materials that do not have a extremely small lattice mismatch between the two unit cell lattice constants, if the twist angle deviation is limited to a small sub-degree scale as can be achieved in experiments \cite{lau2022reproducibility}, the shape of the superlattice will not be distorted too much from a perfect square. For the case of near 90$^\circ$ twisted GeSe, the corresponding moir\'e superlattice structures in real space were depicted in ~\ref{Fig7}(d-f). As shown in these figures, the moir\'e pattern exhibits minimal variation for twist angles of 90.1$^\circ$, 90.2$^\circ$, and 90.5$^\circ$, indicating a high degree of structural stability across these slight angular deviations.

To evaluate the effect of angle deviation on electronic properties, we perform additional calculations for these systems.  To account for small twist angle deviations without being constrained by the commensurate condition, we use two large flake of GeSe stacked at a twist angle close to 90$^\circ$ in real space for each system. The size of each flake is larger than 24 nm, which is sufficient to mimic the bulk. Since each system contains 21952 atoms, density functional theory (DFT) methods become impractical; thus, we employed deep learning-assisted electronic structure calculations (see Appendix for details). The effectiveness of this approach is validated by the band structure calculation of the periodically twisted 90$^\circ$ GeSe bilayer presented in Fig. S3 in the supporting materials. The calculated densities of states (DOS) for structures with slight deviations from 90$^\circ$ are shown in ~\ref{Fig7}(c). These results indicate that the position and the width of the conduction band minimum (CBM) peak in the DOS, corresponding to the flat band, remain nearly unchanged when the twist angle deviates from 90$^\circ$ by up to 0.2$^\circ$, suggesting that small angle deviations have minimal impact on the electronic structure.

\begin{figure}[h]
\includegraphics[width=3.5in]{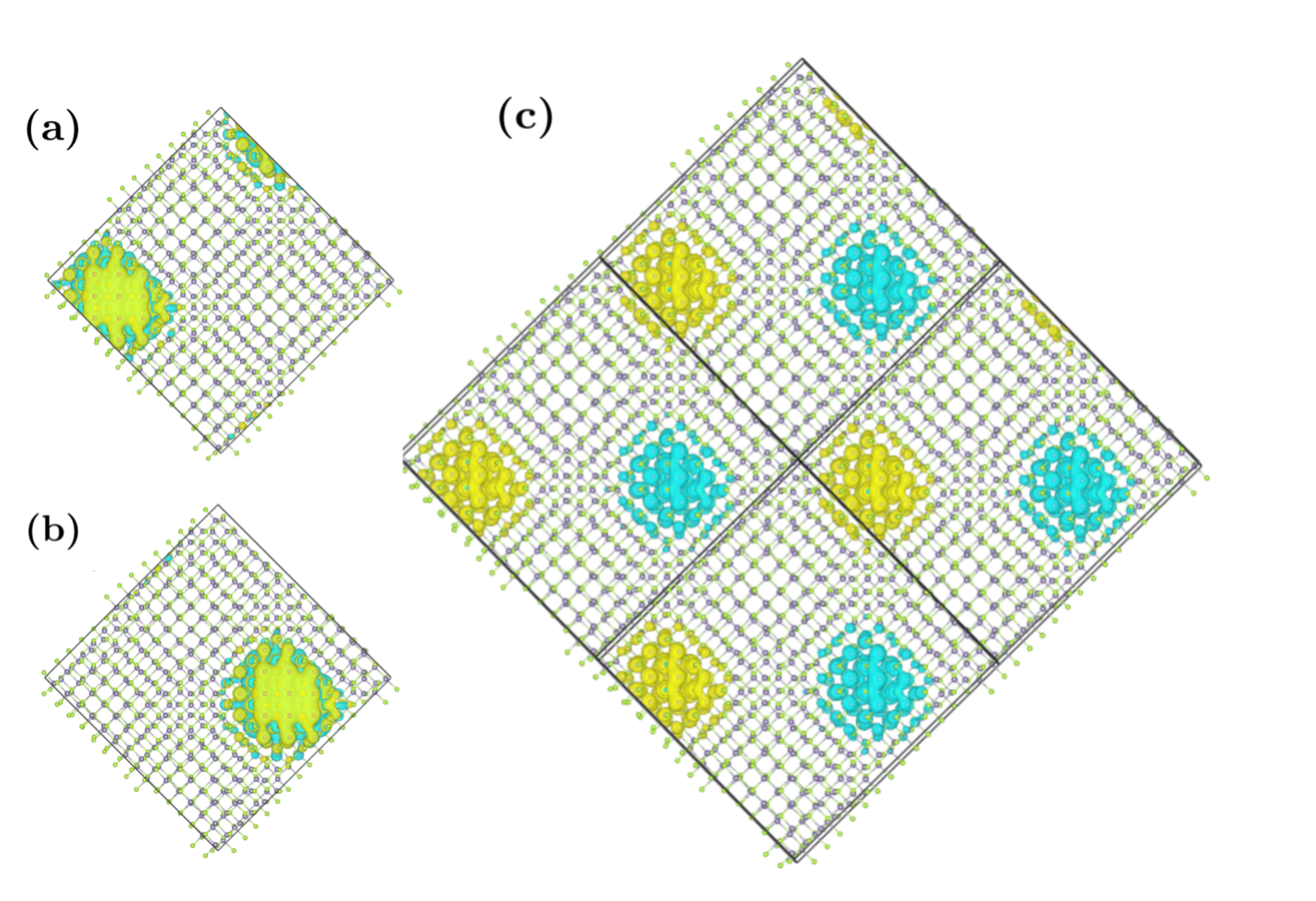}
\centering
\caption{ \captiontitle{Electronic ground states of the moir\'e flat bands in the 90$^\circ$ twisted bilayer GeSe with correlations. } (a-b) Wannier states corresponding to the two flat bands of a checkerboard $2\times1$ supercell of twisted bilayer GeSe. (c) The corresponding magnetization density obtained from DFT$+U+V$ at half-filling.
}
\label{Fig8}
\end{figure}


\subsection{\textbf{\af{Correlation-driven antiferromagentic order in 90$^\circ$ twisted GeX/SnX (X=S, Se) moir\'e superlattices}}}

In flat bands, electronic interactions can dictate electronic properties and give rise to emergent phases in a highly-tunable setting. In order to gain insight in the role of electronic correlations, we complemented our DFT simulations by DFT$+U+V$ simulations, in which an on-site effective Hubbard $U$ and an intersite interaction $V$ (for the localized Wannier orbitals constructed from the DFT flat bands) are added to the DFT Hamiltonian. These effective electronic parameters are evaluated \textit{ab initio} (see the Appendix for more details) in order to determine the ground state of the system when static correlation effects are included.
To this end, we performed several DFT$+U+V$ calculations at half filling of the flat band in the unit cell of twisted GeSe, as well as in doubled $2\times1$ supercells, in order to determine the lowest-energy configuration and allow for collinear and N\'eel antiferromagnetic order. As expected for a square-lattice Hubbard model with weak frustration, we found that the system exhibits an antiferromagnetic instability at half filling \cite{PhysRevB.80.075116}, with the N\'eel antiferromagnetic order been lower than the collinear antiferromagnetic order. However, in contrast to conventional antiferromagnetic insulators composed of atomic orbitals, the effective magnetic moments extend over the charge accumulation regions in the moir\'e unit cell. Each localized Wannier state is found to host a finite magnetization. The Wannier states obtained at the DFT level for the $2\times1$ supercell with N\'eel antiferromagnetic order of GeSe are shown in Figure~\ref{Fig8}(a) and~\ref{Fig8}(b) and the corresponding magnetization, obtained once on-site and intersite interactions are included is shown in Figure~\ref{Fig8}(c). The obtained effective electronic parameters give an on-site interaction $U_{\rm eff} \approx 1.45$meV, corresponding to $U/t \simeq 7.8$, and a ratio of the inter-site interaction to the on-site effective one to be $U_{\mathrm{eff}}/V\approx  10$.

\subsection{\textbf{\af{Realizing $d$-wave superconductivity via 2D heterostructure engineering}}}

Having established a tunable moir\'e realization of a square-lattice Hubbard antiferromagnet at half filling, gate-induced doping and materials compositions now open up new routes for testing longstanding questions regarding the phase diagram of square lattice Hubbard models, mechanisms for unconventional superconductivity, intertwined charge orders and the role of frustration \cite{RevModPhys.66.763,doi:10.1146/annurev-conmatphys-031620-102024,doi:10.1146/annurev-conmatphys-090921-033948}. For strong interactions, prior theoretical studies of the doped square lattice Hubbard model indicate a close competition between charge/spin stripe order and a uniform $d$-wave superconducting phase, which give way to a Fermi liquid at dilute filling \cite{doi:10.1146/annurev-conmatphys-031620-102024}. However, more exotic states such as pair-density wave order have been proposed as well \cite{doi:10.1146/annurev-conmatphys-031119-050711,PhysRevLett.99.127003,PhysRevX.4.031017,doi:10.1146/annurev-conmatphys-090921-033948}.
Electronic frustration via next-nearest-neighbor hopping terms $t'$ plays a key role. While the unfrustrated doped Hubbard model at strong coupling with solely nearest-neighbor hoppings was found by multiple numerical methods to favor stripe order without superconductivity \cite{doi:10.1126/science.aam7127,huang2018,qin2020absence} near optimal doping, prior theoretical work indicates that a finite next-nearest-hopping $t'$ can stabilize $d$-wave superconductivity \cite{PhysRevB.97.045138,10.21468/SciPostPhys.7.2.021,PhysRevB.98.205132,huang2018,PhysRevB.100.195141,doi:10.1126/science.adh7691,jiang2020ground}.

\begin{figure}[h]
\includegraphics[width=0.9\columnwidth]{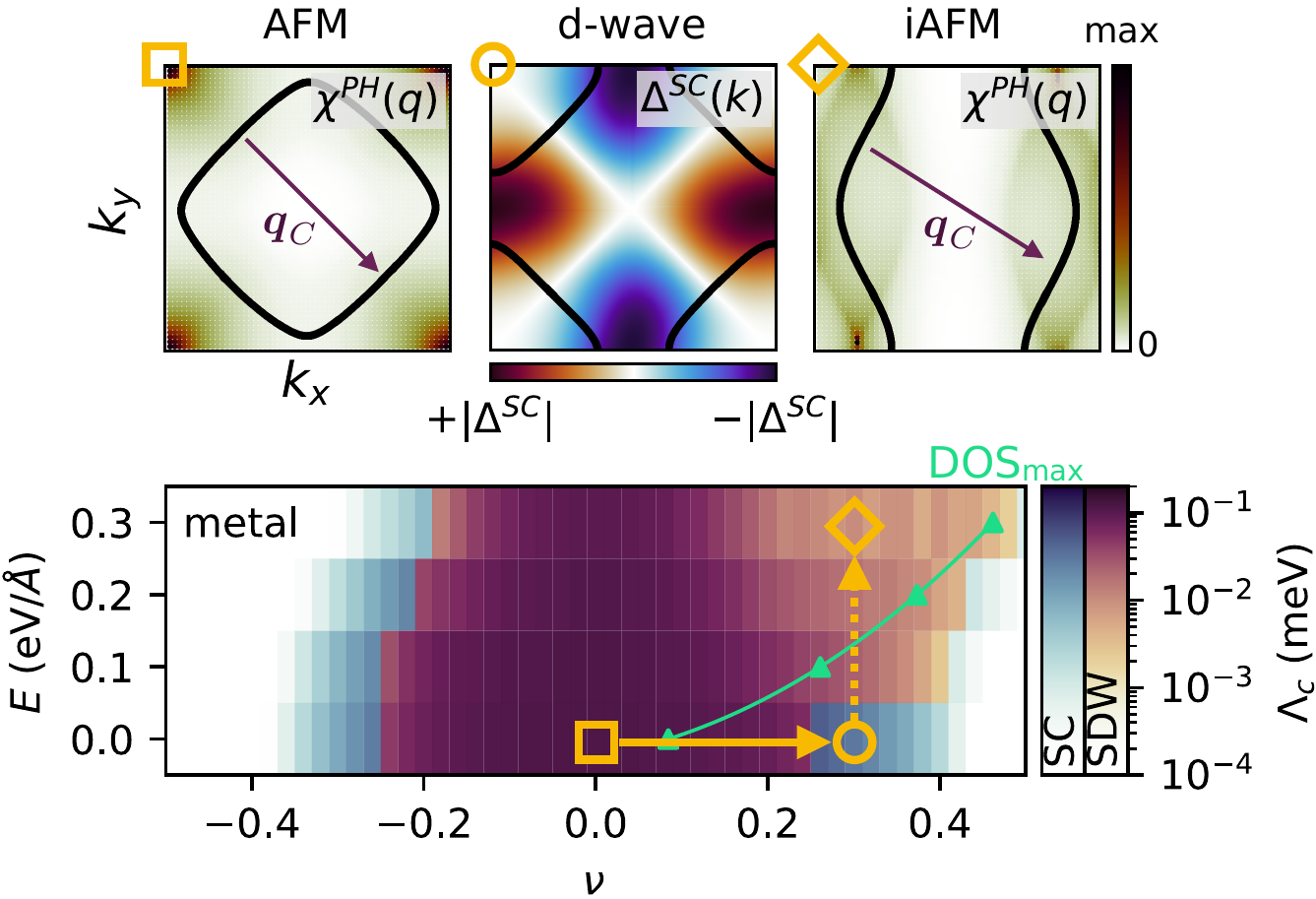}
\centering
\caption{ 
\af{
\captiontitle{FRG phase diagram of $90^{\circ}$ twisted GeSe in the presence of an external displacement field $E$}. 
The phase diagram shows the critical scale $\Lambda_{\mathrm c} \sim T_{\mathrm{c}}$ of the leading Fermi surface instability and indicates regimes of spin density wave order (SDW, red), superconductivity (SC, blue) and un-ordered metallic regimes.
At half-filling ($\nu=0$), the system features antiferromagnetic order (yellow square) and the particle-hole susceptibility $\chi^{\text{PH}}(\bvec q)$ is peaked at the commensurate nesting vector $\bvec q_C = (\pi, \pi)$. 
Upon electron (hole) doping, the nesting conditions are weakened such that AFM fluctuations give rise to $d$-wave superconductivity (yellow circle).
In the presence of an external displacement field, the Fermi surface becomes anisotropic and is stretched in $\hat y$-direction, which weakens superconductivity and strengthens incommensurate AFM order (iAFM, yellow diamond), whose anisotropy axis follows the leading bosonic momentum transfer $\bvec q_C$.
The green line indicates the van-Hove singularity on the electron-doped side that tunes to larger electron densities as the displacement field is increased.
%
}
}
\label{fig:frg}
\end{figure}

\af{
Here, we elucidate this question by addressing electronic correlations by unbiased functional renormalization group (FRG)~\cite{metznerFunctionalRenormalizationGroup2012,platt2013functional} calculations of twisted bilayer GeSe starting from the \textit{ab initio} electronic parameters.
The FRG smoothly connects the non-interacting model to an effective low-energy theory by successively integrating out high-energy degrees of freedom above a flowing energy cutoff $\Lambda = \infty \to 0$ such that quantum fluctuations are systematically included in the effective two-particle interaction vertex.
By perturbatively expanding possible scattering processes, the FRG allows for an unbiased treatment of electronic order and symmetry-breaking transitions in the pairing, charge and magnetic channel. A divergent interaction in one of the channels at scale $\Lambda_{\mathrm{c}}\sim T_{\mathrm{c}}$ indicates the transition to a symmetry-broken electronic phase, see Method Section for details.
This renders FRG a suitable method to analyze the competition of antiferromagnetic order and superconductivity in the frustrated $t-t'$ low-energy models of 90$^\circ$ twisted GeX/SnX (X=S, Se) moir\'e superlattices.

The results of the FRG study are summarized in Fig.~\ref{fig:frg}. The phase diagram shows the critical scale $\Lambda_{\mathrm c}$ of the leading Fermi surface instability and indicates regimes of spin density wave order (SDW, red), superconductivity (SC, blue) and metallic regimes without electronic order.
At half-filling ($\nu=0$, yellow square), the FRG predicts an extended range of antiferromagnetic (AFM) order that is driven by nesting between the approximate $C_{4v}$-symmetric Fermi surface sheets with vector $\bvec q_C = (\pi, \pi)$ in line with the DFT+$U$+$V$ analysis provided in the previous paragraph.
The particle-hole susceptibility $\chi^{\text{PH}}(\bvec q)$ in the AFM phase and the respective Fermi surface (black line) are shown in the upper panel of Fig.~\ref{fig:frg}.
Upon electron (hole) doping, the nesting conditions are weakened such that AFM order eventually gives way to $d$-wave superconductivity at $\nu \sim 0.3$ (yellow circle), driven by fluctuations of the nearby AFM state.
The superconducting order parameter $\Delta^{\text{SC}}(\bvec k)$ is shown in the upper panel of Fig.~\ref{fig:frg} and exhibits gapless points whenever the Fermi surface intersects with zeros of the order parameter.
}

\af{While we exemplified the emergence of $d$-wave superconductivity in twisted bilayer of GeSe, we stress the generality of our approach suggesting similar correlated states to emerge in the family of other 90$^\circ$ twisted GeX/SnX (X=S, Se) moir\'e superlattices.
}
To this end, the materials parameterized in Table~\ref{tab:table2} provide a route to simulate the role of frustration in the square lattice Hubbard model in an experimental setting via heterostructure composition. Notably, four-layer twisted GeSe configurations most closely mirror weak frustration $|t'/t| \approx 0.2$ representative of cuprate high-temperature superconductors \cite{RevModPhys.78.17,doi:10.1126/science.aal5304}. In contrast to cuprates, electron and hole doping away from half filling are flipped by virtue of $t'$ and $t$ carrying the same sign. 

\subsection{\textbf{\af{Simulating the competition of stripe order vs. $d$-wave superconductivity via external displacement fields}}}

Interestingly, the proposed moir\'e realization of a two-dimensional Hubbard model can also serve as a simulator to explore competing ground states at finite filling. For instance, while the unfrustrated Hubbard model at 1/8 doping was found by multiple methods to favor a period-8 stripe phase without superconductivity \cite{doi:10.1126/science.aam7127,huang2018}, the stability of stripe order, variations of its period, and the onset of $d$-wave superconductivity for variations of either filling or frustration has attracted much attention \cite{PhysRevB.97.045138,10.21468/SciPostPhys.7.2.021,PhysRevB.98.205132,huang2018,PhysRevB.100.195141}. 

\begin{figure}[h]
\centering
\includegraphics [width=3.5in] {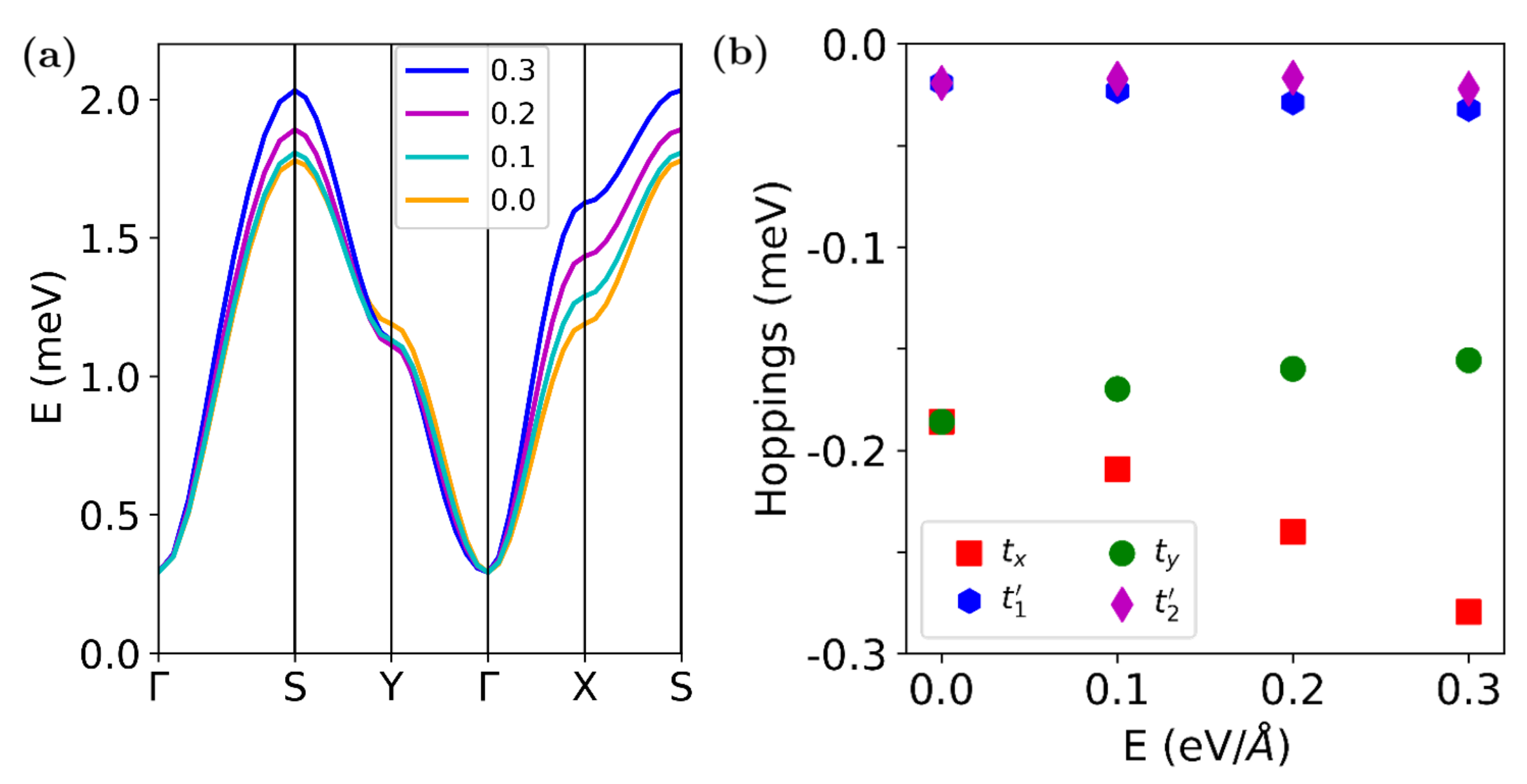}
\caption{\captiontitle{ Electric field tuning of moir\'e flat bands in 90$^\circ$ twisted bilayer GeSe.} (a) The flat bands at the CBE under various vertical electric field ($E$) with the value of $0.0$, $0.1$, $0.2$ and $0.3$ eV/\AA. (b) The effective model fitting parameters as a function of the electric field.} 
\label{Fig9}
\end{figure}

\af{We propose that our design principle for 90$^\circ$ twisted GeX/SnX (X=S, Se) moir\'e superlattices allows to investigate this question by tuning the anisotropy of the Fermi surface in-situ via the application of external displacement fields -- favoring either striped-AFM order or $d$-wave superconductivity.
Figure~\ref{Fig9}~(a) shows the low-energy flat bands of the CBE of twisted bilayer GeSe for four different strengths of the external electric field of $0.0$, $0.1$, $0.2$ and $0.3$ eV/\AA. 
For finite displacement field, the in-plane $C_{2d}$ rotational symmetry is broken, leading to an energetic offset between the $X$-point and the $Y$-point of the moir\'e Brillouin zone.
The strength of the anisotropy is steered by the strength of the displacement field, while the overall bandwidth of the moir\'e flat band is only slightly altered.
We summarize the effective hoppings parameters for the anisotropic $t-t'$ models of twisted bilayer GeSe in Table~\ref{tab:table2} and show their displacement field dependence in Fig.~\ref{Fig9}~(b). 
For experimentally relevant field strengths, the effective hopping amplitudes can be tuned over a wide range, highlighting the flexibility of our engineering approach in tuning electronic properties in 90$^\circ$ twisted moir\'e superlattice.
}

\af{
The electronic structure engineering also has profound impact on the correlated electronic phases and the competition between striped AFM order and $d$-wave superconductivity as shown in Fig.~\ref{fig:frg}.
In the presence of an external displacement field, the Fermi surface becomes anisotropic and is stretched in $\hat y$-direction (yellow diamond).
This weakens $d$-wave superconductivity and strengthens incommensurate AFM order (iAFM). Maxima of the particle-hole susceptibility $\chi^{\text{PH}}(\bvec q)$ in this phase are shifted away from $(\pi, \pi)$, which indicates the formation of striped-AFM order whose anisotropy axis follows the leading momentum transfer $\bvec q_C$. 
To this end, the formation of striped AFM order on the electron-doped side is facilitated by the emergence of a tunable van-Hove singularity on the electron-doped side (green line) that tunes to larger electron densities as the displacement field is increased.
Regions of superconducting order are instead displaced to larger electron (hole) doping, however, survive even in the presence of anisotropies at larger electron (hole) dopings.
Therefore, we argue that 90$^\circ$ twisted GeX/SnX (X=S, Se) moir\'e superlattices provide an ideal testbed to tune between striped AFM order and $d$-wave superconductivity via displacement field control of the anisotropy in the system.
}

\begin{figure}[h]
\centering
\includegraphics [width=3.5in] {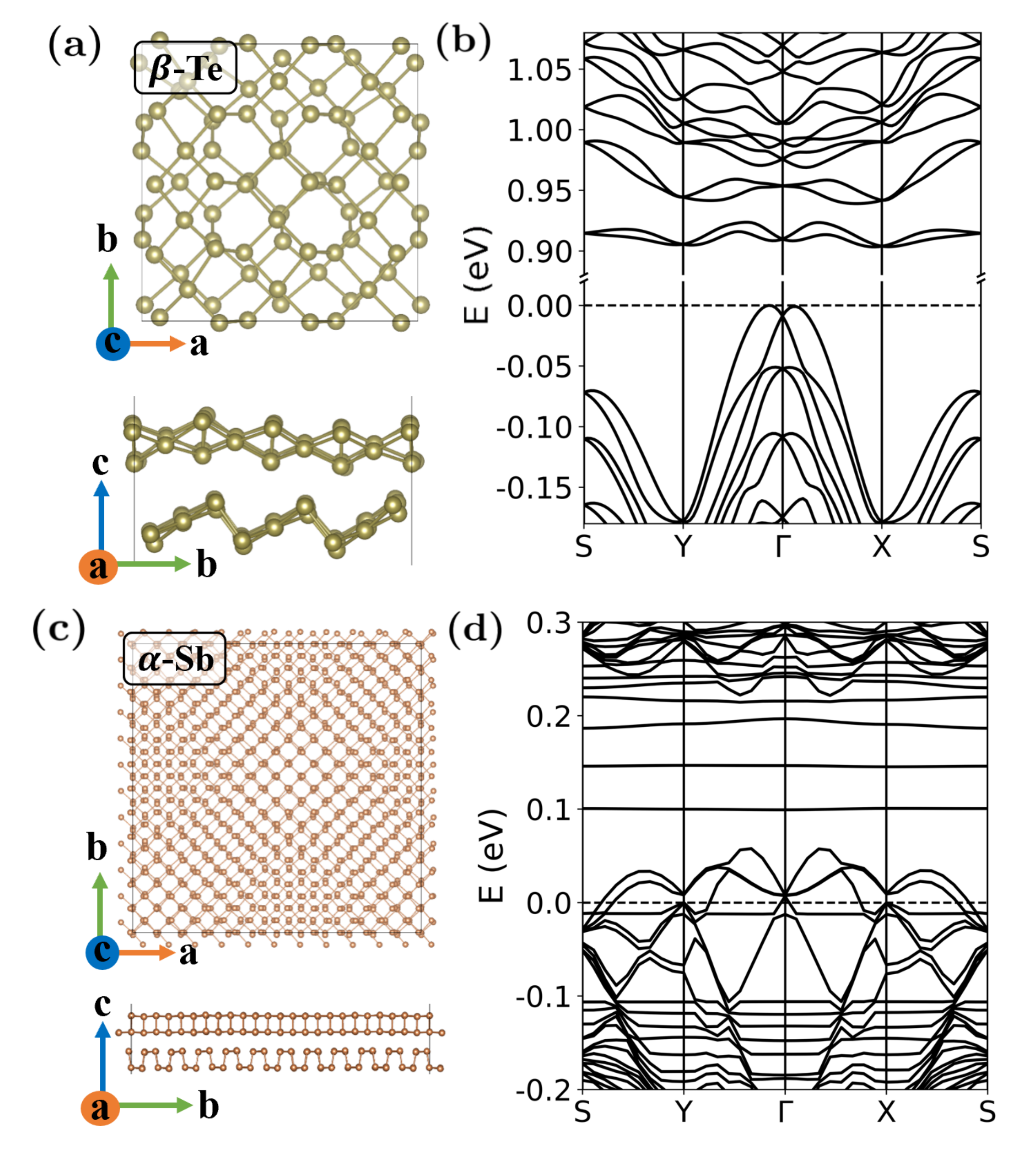}
\caption{The atomic(a, c) and the electronic structures (b, d) of 90$^\circ$ twisted bilayer $\beta$-Te and $\alpha$-Sb, respectively. The dashed line represents the Fermi level.}
\label{Fig10}
\end{figure}

\subsection{\textbf{Other potential material candidates}} 
 The general strategy we propose for constructing moir\'e square superlattices can also be applied to other 2D materials with rectangular lattices. We specifically investigate moir\'e square superlattices formed by twisting $\beta$-tellurene and $\alpha$-antimonene. Both $\beta$-tellurene and $\alpha$-antimonene have been successfully synthesized or exfoliated down to monolayers in experiments \cite{wang2018field,shi2019van,shi2020tuning}, making them viable candidates for fabricating 90$^\circ$ twisted bilayer or multilayer systems. The atomic structures and calculated band structures of 90$^\circ$ twisted bilayer $\beta$-tellurene and $\alpha$-antimonene are shown in Figure~\ref{Fig10}. The moir\'e square lattices of these systems have lattice constants of 17 nm and 52 nm, respectively. As shown in Figures~\ref{Fig10} (b) and (d), isolated flat bands emerge at the band edges of both systems, with bandwidths of less than 20 meV for $\beta$-tellurene and 1.5 meV for $\alpha$-antimonene. These properties suggest that these systems could be utilized to explore Hubbard model physics on a square lattice. 
 In addition to the materials presented in this study, we identify more than 40 other potential 2D material candidates with rectangular lattices, listed in Table~\ref{tab:table3}. These materials have exfoliation energies comparable to that of black phosphorene, making them suitable for exfoliation from bulk to fabricate 90$^\circ$ twisted layers. Furthermore, they possess lattice constants that result in moir\'e superlattice wavelengths between 2 nm and 10 nm. This range is ideal for the formation of flat bands while maintaining structural homogeneity in experimental systems.

\section{\textbf{Conclusions}}
Our work proposes a general scheme to use rectangular-lattice 2D van der Waals materials to construct moir\'e square superlattices. We predict that moir\'e heterostructures of 90$^\circ$ twisted bilayers and multilayers of the rectangular-lattice transition-metal dichalcogenide family (Ge/Sn)(S/Se) can provide a platform to realize the paradigmatic square-lattice Hubbard model with repulsive interactions in a tunable setting. In this set of compounds, layer composition grants a twofold handle to tune electronic frustration and drive the material into a regime with narrow square-lattice moir\'e bands dominated by strong Coulomb interactions. Focusing on 90$^\circ$ rotated bilayers of GeSe, we computed the strength of Coulomb interactions and predict the formation of a paradigmatic square-lattice Mott insulator with N\'eel order at half filling. 
Combined with gate-tunable carrier doping, this suggests that 90$^\circ$-twisted (Ge/Sn)(S/Se) can serve as an experimental solid-state platform to simulate the phase diagram of the copper-oxygen planes of cuprate high-temperature superconductors, to provide new experimental insight into the nature of the pseudogap, the role of electronic frustration, and competing and intertwined superconducting and density wave orders. 

{\it Note added:} After completing this work we came across
a related paper by P. Myles Eugenio and collaborators \cite{eugenio2024tunable}. Their work proposes realizing a similar square lattice Hubbard model with frustration (or the $t-t'-U$ model) using twisted square homobilayer, based on their model only calculations.

\acknowledgements
L.X. and Q.X. acknowledge the support by the National Key Research and Development Program of China (Grant No. 2022YFA1403501), Guangdong Basic and Applied Basic Research Foundation (Grant No. 2022B1515120020), the National Natural Science Foundation of China (Grant No. 12504218 and No. 62341404), the Hefei National Research Center for Physical Sciences at the Microscale (KF2021003), and the Max Planck Partner group programme. D.M.K acknowledge funding by the Deutsche Forschungsgemeinschaft (DFG, German Research Foundation) within the Priority Program SPP 2244 “2DMP” - 443274199 and under Germany’s Excellence Strategy - Cluster of Excellence Matter and Light for Quantum Computing (ML4Q) EXC 2004/1 - 390534769. M.C. acknowledges support by the U.S. Department of Energy (DOE), Office of Basic Energy Sciences, under Award No. DE-SC0024494. We acknowledge computational resources provided by RWTH Aachen University under project number rwth0763. This work was supported by the Max Planck-New York City Center for Nonequilibrium Quantum Phenomena. This work was supported by the European Research Council (ERC-2015-AdG694097), the Cluster of Excellence ‘Advanced Imaging of Matter' (AIM) and Deutsche Forschungsgemeinschaft (DFG) -SFB-925–project 170620586.

\appendix

\section{\textbf{MODEL AND COMPUTATIONAL APPROACHES}}
{\it Density functional theory calculations} --- The present calculations are done within the framework of density functional theory (DFT) using the plane-wave pseudopotential approach performed in the VASP code \cite{kresse1996efficient}.
The electron-ion interactions can be described by the projector augmented wave (PAW) pseudopotentials \cite{blochl1994projector}. Configurations of 4$\emph{s}^2$4$\emph{p}^2$ for Ge, 5$\emph{s}^2$5$\emph{p}^2$ for Sn, 3$\emph{s}^2$3$\emph{p}^4$ for S and 4$\emph{s}^2$4$\emph{p}^4$ for Se are considered as valence electrons. We use the Perdew-Burke-Ernzerhof (PBE) parameterization of the generalized gradient approximation \cite{perdew1996generalized} as the exchange correlation functional. We calculate the equilibrium lattice parameters of GeS in the bulk phase and found that the DFT-TS vdW functionals \cite{tkatchenko2009accurate} provide better agreement with the experimental values \cite{wiedemeier1978refinement} within less than 1$\%$ errors. The DFT-TS vdW functional is then adopted for all calculations. A plane-wave basis set with an energy cutoff of 400 eV and a 1$\times$1$\times$1 momentum grid are used to characterize the ground state and the mechanical relaxation.
Our calculations are fully relaxed (w.r.t.~all the atoms), which is known to be important in other moir\'e systems to avoid artificial effects stemming from unrelaxed structures. 
The relaxation procedure ensures that the force on each atom converges to values smaller than $0.01~\mathrm{eV}/\AA$.
A long z-direction auxiliary vacuum region larger than $15~\AA$ is added to avoid the artificial interaction between the periodic slabs. As spin-orbit coupling (SOC) is found to have negligible effects on the flat bands calculated for the systems, we exclude SOC in the calculations.

\begin{table}[h]
    \centering
    \captionsetup{position=below,justification=raggedright,font=footnotesize}
    \caption{\captiontitle{Tight-binding parameters of the moir\'e flat bands for 90$^\circ$ twisted multilayer GeX/SnX fitted to DFT bandstructures.}}
    \label{tab:table2}
    \resizebox{0.48\textwidth}{!}{  
      \begin{tabular}{|c||c|c|c|c|c|c|}
        \hline\hline
        Materials & 
        \begin{tabular}[c]{@{}c@{}}$t_x$\\(meV)\end{tabular} & 
        \begin{tabular}[c]{@{}c@{}}$t_y$\\(meV)\end{tabular} & 
        \begin{tabular}[c]{@{}c@{}}$t^{\prime}_1$\\(meV)\end{tabular} & 
        \begin{tabular}[c]{@{}c@{}}$t^{\prime}_2$\\(meV)\end{tabular} &
        \begin{tabular}[c]{@{}c@{}}$t^{\prime}_1$/$t_x$\\(\%)\end{tabular}  &
        \begin{tabular}[c]{@{}c@{}}$t^{\prime}_2$/$t_y$\\(\%)\end{tabular}  \\
        \hline\hline
        2L-GeS &
           -$6.054 $ & -$6.054$ & $0.320$ & $0.290$ & -$5.2$  & -$4.8$\\
        2L-SnS &
           -$0.410$  & -$0.410$ & -$0.016$ &-$0.012$ & $3.9$  & $2.9$ \\
        2L-GeSe &
           -$0.186$  & -$0.186$ & -$0.010$ & -$0.027$ & $5.3$  & $14.5$\\
        2L-GeSe (0.1)$^{*}$ &
           -$0.209$  & -$0.169$ & -$0.012$ & -$0.029$ & $5.7$  & $17.1$\\
        2L-GeSe (0.2)$^{*}$ &
           -$0.240$  & -$0.158$ &  -$0.014$ &-$0.032$ & $5.8$  & $20.2$\\
        2L-GeSe (0.3)$^{*}$ &
           -$0.281$ &-$0.153$  & -$0.018$ &-$0.037$ & $6.4$ & $24.1$\\
        3L-GeSe &
           -$0.481$ & $0.527$   & -$0.010$ &-$0.040$ & 2.1  &  -7.6\\
        4L (III-I)-GeSe &
           -$0.947$ & $0.967$   & -$0.010$ &-$0.081$ & 1.1 & -8.4\\
        4L (II-II (AB-A$^t$B$^t$))-GeSe &
           $0.188$ & $0.172$    & -$0.050$ &-$0.028$ & -26.6 & -16.3\\ 
        4L (II-II (AB-B$^t$A$^t$))-GeSe &
           0.199 & 0.199    & -0.036 &-0.040 & -18.1 & -20.1 \\ 
        \hline\hline
      \end{tabular}
    }
  \footnotesize{*The asterisk denotes results with external electric field. The value of the applied field is indicated in the parentheses with unit of eV/\AA}.
\end{table}

{\it Deep-learning assisted electronic structure calculation} --- The electronic structure calculation for the non-periodic moir\'e supercell slab has been encapsulated in the DeepH package \cite{li2022deep}. The key idea is to employ machine learning to study the relationship between the Hamiltonian matrix and material structure, and then to make predictions for new systems. By constructing hundreds of small structures, a corresponding database of Hamiltonian is obtained using DFT methods and subsequently trained by neural networks. Thereafter, trained network models can be used to predict the band structure of larger structures. To train the DeepH model for moir\'e twisted GeSe bilayers, we prepared datasets (625 configurations) from zero-twist-angle $5\times5\times1$ bilayer supercells by shifting the top layer within the 2D plane and inserting random perturbations to each atomic site, based on the relaxed primitive GeSe bilayer. Adopting the same constructing manner, we have also built a group (82 configurations) of unrelaxed 90$^\circ$ twisted bilayer GeSe. All these dataset structures were calculated using localized pseudoatomic orbitals (PAOs) and pseudopotentials as implemented in the OPENMX code \cite{ozaki_openmx}, due to the localization ability of the basis. The exchange-correlation functional, treated by the generalized gradient approximation of the Perdew-Burke-Ernzerhof (PBE) form \cite{perdew1996generalized}, and norm-conserving pseudopotentials \cite{morrison1993nonlocal}, and the DFT-D3 method corrected van der Waals interaction \cite{grimme2010consistent} were used in the calculations. The basis sets for the two elements in the system were chosen as Ge7.0-$\emph{s}^3$$\emph{p}^2$$\emph{d}^2$ (RC = 7.0 Bohr) and Se7.0-$\emph{s}^3\emph{p}^2\emph{d}^2$, which include 19 atomic-like basis functions. The energy cutoff was set to 300 Ry in real space for numerical integration and solving the Poisson equation. A $7\times6\times1$ and $2\times2\times1$ mesh of k-points were adopted in preparing the datasets of $5\times5\times1$ GeSe bilayer and 90$^\circ$ twisted bilayer GeSe, respectively. The convergence criterion was set as a total energy difference of less than $4\times10^{-7}$ Hartree.
During the training step, a message-passing neural network (MPNN) implemented in the PyTorch-Geometric Python library \cite{fey2019fast} was used to represent the Hamiltonian matrix, where the vertices, edges, and edge embeddings represent atoms and atom pairs, and Hamiltonian matrix blocks. The MPNN network includes five message-passing (MP) layers and one local coordinate message-passing (LCMP) layer. The batch size was fixed at 2, meaning that each batch contained two material structures. The loss function was set to the default value, and the learning rate was initially set to 0.001. The total training epoch was 6000. By providing the overlap matrix of the large system from the revised OPENMX code \cite{ozaki_openmx} and the predicted Hamiltonian from the DeepH code, the eigenvalues of the large system can be obtained by solving the eigenvalue problem \cite{wang2019first}.

\begin{table}[htb!]
\centering
    \caption{\captiontitle{Other potential 2D materials candidates for realizing square lattice Hubbard model in 90$^\circ$ twisted systems.} (a, b) denote the two lattice constants in the primitive unit cell; E$_b$ represents the binding energy calculated using rVV10 functionals from MC2D database \cite{mounet2018two}; E$_g$ represents the band gap obtained from PBE functionals; the moir\'e period L is estimated by L=[(1/a)-(1/b)]$^{-1}$; and the ``3D parent ID" column lists the identification numbers of the corresponding 3D parent compounds for each 2D layered material.} 
    \label{tab:table3}
\resizebox{0.48\textwidth}{!}{
    \begin{tabular}{|c||c|c|c|c|c|} 
    \hline\hline
    Materials & 
    \begin{tabular}[c]{@{}c@{}} Unit cell(a,b)\\/\AA\end{tabular} &
    \begin{tabular}[c]{@{}c@{}} E$_b$\\/meV/\AA$^2$\end{tabular} & 
    \begin{tabular}[c]{@{}c@{}} E$_g$\\/eV\end{tabular} &
    \begin{tabular}[c]{@{}c@{}} L \\/\AA\end{tabular} & 
    3D parent ID  \\ 
    \hline\hline
    SnSe     & (4.15, 4.44)   & 35.72 & 0.96 & 63.53 & COD (9008786)  \\
    GeTe     & (4.22, 4.37)   & /     & 0.80 & 122.00   & ICSD (638005)  \\
    $\beta$-PbO      & (4.87, 5.55)   & 27.20 & 2.53 & 39.62 & COD (2310433)  \\ 
    SnO      & (4.53, 5.31)   & 28.40 & 2.10 & 30.83 & ICSD (424729)  \\
    MoO$_3$     & (3.71, 3.87)   & 26.64 & 1.53 & 87.08 & ICSD (80577)   \\ 
    PdSeO$_3$  & (6.53, 7.27)   & 29.70 & 0.40 & 64.15 & ICSD (415955)  \\
    HfNBr    & (3.57, 4.12)   & 18.17 & 2.13 & 26.62 & COD (1532008)  \\ 
    ZrNI     & (3.75, 4.14)   & 20.20 & 1.28 & 39.57 & ICSD (36119)   \\
    ZrNBr    & (3.62, 4.14)   & 18.29 & 1.80 & 28.45 & ICSD (27393)   \\ 
    TeO$_2$   & (5.32, 5.74)   & 54.8 & 2.30 & 72.70 & COD (1011183)   \\
    PdI$_2$     & (6.9, 8.36)    & 33.17 & 1.05 & 39.38 & ICSD (25120)   \\ 
    PtI$_2$     & (6.88, 8.52)   & 31.88 & 1.35 & 35.63 & ICSD (60760)   \\ 
    ScOBr    & (3.57, 3.98)   & 17.28 & 3.07 & 34.43 & COD (2206575)  \\
    InOBr    & (3.64, 4.14)   & 17.30 & 2.26 & 30.41 & ICSD (24059)   \\ 
    InOCl    & (3.56, 4.17)   & / & 2.59 & 24.33 & ICSD (24058)   \\
    InTeBr   & (7.56, 8.39)   & 19.65 & 2.26 & 76.87 & COD (1524712)  \\ 
    InTeI    & (7.77, 8.63)   & 19.03 & 1.73 & 78.00 & COD (1525202)  \\ 
    SbOCl    & (9.44, 10.94)  & 45.48 & 2.05 & 68.64 & ICSD (24751)   \\
    TiNBr    & (3.36, 3.95)  & 20.10  & 0.60 & 22.49  & ICSD (27395) \\
    NbO$_2$I    & (3.77, 3.91)   & 21.53 & 0.81 & 109.46 & ICSD (418061)   \\
    Nb$_2$SiTe$_4$ & (6.34, 7.87)   & 28.10 & 0.56 & 32.66 & ICSD (81416)   \\ 
    Nb$_2$GeTe$_4$ & (6.47, 7.9)    & 27.47 & 0.47 & 35.76 & ICSD (81417)   \\ 
    ZnAs$_2$O$_4$  & (4.54, 4.99)   & 27.98 & 3.39 & 49.56 & COD (9001071)  \\ 
    W$_4$PBr$_{11}$  & (10.45, 11.93) & 18.45 & 0.56 & 83.87 & COD (4341586)  \\ 
    W$_4$PCl$_{11}$  & (9.94, 11.42)  & 17.86 & 0.57 & 77.03 & ICSD (422269)  \\ 
    Re$_2$S$_2$O$_{13}$ & (8.38, 9.77)   & 18.25 & 2.10 & 58.85 & COD (4324293)  \\
    LiBiO$_2$   & (4.73, 5.27)   & 38.68 & 2.27 & 46.66 & ICSD (25385)   \\ 
    KAs$_2$F$_7$   & (7.21, 8.41)   & 31.27 & 4.19 & 50.63 & ICSD (36332)   \\ 
    Li$_2$ZnBr$_4$ & (6.5, 7.73)    & 44.93 & 3.47 & 40.87 & COD (1517836)  \\ 
    Nb$_2$PbO$_6$  & (4.81, 5.28)   & 21.89 & 1.89 & 53.93 & COD (2101515)  \\ 
    AlBiCl$_6$  & (11.89, 17.92) & 16.52 & 3.59 & 35.36 & ICSD (414261)  \\  
    CdP$_2$O$_{10}$  & (8.63, 10.22)  & 27.94 & 1.36 & 55.39 & COD (9015952)  \\ 
    Hg$_2$TeBr$_3$ & (10.22, 13.47) & 31.49 & 1.66 & 42.39 & ICSD (417407)  \\ 
    Na$_2$ZnCl$_4$ & (6.4, 8.05)    & 41.28 & 4.16 & 31.10 & ICSD (402063)  \\ 
    Na$_2$PSe$_3$  & (8.0, 11.76)   & 35.33 & 1.75 & 25.05 & ICSD (415240)  \\ 
    NaNbCl$_6$  & (6.33, 6.82)   & 18.85 & 2.21 & 88.92 & ICSD (36518)   \\ 
    NaGaH$_4$   & (6.37, 6.98)   & 25.36 & 4.44 & 73.04 & ICSD (67309)   \\ 
    KAuS     & (6.53, 7.85)   & 42.52 & 1.68 & 38.91 & COD (1510204)  \\ 
    KAuSe    & (6.85, 8.1)    & 39.08 & 1.49 & 44.29 & COD (1510205)  \\ 
    KC$_2$N$_3$    & (6.93, 8.43)   & 32.53 & 4.37 & 38.88 & ICSD (411930)  \\ 
    KO$_3$Cl    & (4.65, 5.57)   & 23.89 & 5.40 & 28.09 & COD (2310654)  \\ 
    KCN$_7$     & (6.39, 8.09)   & 23.87 & 3.50 & 30.42 & COD (4103810)  \\ 
    K$_2$PdAs$_2$  & (5.94, 6.39)   & 44.36 & 0.54 & 83.71 & ICSD (32009)   \\ 
    KSb$_2$F$_7$   & (7.52, 8.21)   & 25.87 & 3.93 & 88.94 & COD (4343786)  \\ 
    K$_2$SiAs$_2$  & (6.44, 6.98)   & 36.97 & 0.96 & 83.78 & ICSD (40426)   \\ 
    K$_2$SiP$_2$   & (6.09, 6.79)   & 41.24 & 1.19 & 59.21 & ICSD (36367)   \\ 
    K$_2$WSe$_4$   & (9.73, 12.43)  & 38.92 & 1.58 & 44.83 & COD (2005921)  \\ 
    K$_3$InP$_2$   & (6.77, 7.57)   & 37.82 & 0.74 & 63.73 & ICSD (300141)  \\ 
    K$_2$GeSe$_4$  & (8.13, 9.11)   & 26.09 & 1.78 & 75.50 & ICSD (78828)   \\ 
    K$_2$GeTe$_4$  & (8.34, 10.21)  & 24.63 & 0.89 & 45.59 & ICSD (38341)   \\ 
    \hline
    \hline
    \end{tabular}
}
\end{table}  

{\it DFT+U+V calculations} --- The \textit{ab initio} evaluation of the effective electronic 
parameters (on-site effective Hubbard $U_{\mathrm{eff}}=U-J$ and the intersite interaction $V$) and the DFT+U+V
calculations are performed using the Octopus code~\cite{tancogne2020octopus,acbn0}.  A real-space spacing of 0.4 Bohr is
chosen, and we employ HGH norm-conserving pseudopotentials~\cite{hgh_pseudos}. The LDA is used for describing the local DFT part. The vacuum size, atomic coordinates and lattice constant are taken to be the same as for the DFT treatment, as described above. In order to analyse if the system present or not an antiferomagnetic instability when correlations are taken into account, we constructed a doubled-cell by repeating the DFT unit cell along the $x$-axis. We employed a $3\times3$ k-point grid for the unit cell and the $1\times1$  k-point grid for the double cells.
The localized subspace was constructed using the Wannier90 library~\cite{Pizzi2020}, taking the flat bands obtained from the LDA calculation performed with the Octopus code. 
We then populated these bands at half filling by adding an excess charge of two electrons in the system and performed DFT+U+V calculations. We performed non-polarized and spin-polarized calculations for the unit cell, as well as a spin-polarized calculations for the double cells. In all spin-polarized calculations, we started from random magnetic configurations in order to determine the ground-state.
Only the first nearest neighbor interaction was considered in the intersite interaction. 

The standard approach to estimate the on-site UU and intersite VV rely on the cRPA method. This method is however prohibitively expensive for been applied to twisted bilayer systems, and an alternative approach needs to be employed. In order to get a first estimate on the relative strength of the on-site and intersite interaction, we compute first the Coulomb integrals of the bare Coulomb potential.
\begin{equation}
    \langle  \phi_m^I\phi_m^I | v |\phi_{m'}^J\phi_{m'}^J\rangle = \int d\mathbf{r}  d\mathbf{r'} |\phi_m(\mathbf{r})|^2 \frac{1}{|\mathbf{r}-\mathbf{r'}|} |\phi_{m'}(\mathbf{r'})|^2\,,
\end{equation}
With this approach, we get $U_{bare}= 1.99$eV and $V_{bare} = 235$meV, giving a ratio $U_{bare}/V_{bare} = 8.46$. 
While these values might seems to be very large, we can compare them to the estimate we would get for a $1s$ hydrogenic orbital of the form $\phi_{1s}(\mathbf{r}) = \left(\frac{Z^3}{\pi}\right)^{1/2} e^{-Zr}$, the on-site Coulomb integral $\langle  \phi_{1s}\phi_{1s} | v |\phi_{1s}\phi_{1s}\rangle = \frac{5}{8}Z$. If we choose $Z$ such that the spread of the Wannier function is the same as the spread of the Wannier state given by Wannier90 (here $\approx 30$\AA), we obtain that $Z\approx0.25$. This would lead to $\langle  \phi_{1s}\phi_{1s} | v |\phi_{1s}\phi_{1s}\rangle \approx 4.25$eV.

We now proceed with estimating the screened Coulomb interaction.
Key to the two-dimensional materials is the change of screening compare to three-dimensional materials. To take into account the two-dimensional nature of the material, we compute the screened Coulomb integrals for the twisted bilayer system, replacing the bare Coulomb potential by the Rytova-Keldysh potential\cite{rytova2018screened,keldysh2024coulomb} which reads in real-space
\begin{equation}
    V_{2D}(s) = \frac{\pi}{2\rho_0}\Big[ H_0\left(\frac{s}{\rho_0} \right) - Y_0\left(\frac{s}{\rho_0} \right)   \Big]\,,
\end{equation}
where $H_0$ and $Y_0$ are respectively the Struve and Bessel functions of the second kind and $\rho_0$ is a screening length, which is obtained here \textit{ab initio} from a calculation for the untwisted bilayer calculation as explained above. 
In Fourier space, this potential reads
\begin{equation}
    V_{2D}(\mathbf{q}_{\parallel}) = \frac{2\pi}{|\mathbf{q}_{\parallel}|(1+\rho_0|\mathbf{q}_{\parallel}|)}\,,
\end{equation}
ad properly captures the asympotic behavior of the two-dimensional screening\cite{PhysRevB.84.085406}. In order to determine the screening length $\rho_0=\frac{d\epsilon}{2}$, we use a method originally proposed by Qiu et al.\cite{PhysRevB.93.235435} which consists in computing an effective \textit{ab initio} screening from calculations done in a supercell
\begin{equation}
    \epsilon_{2D}^{-1}(q) = \frac{q}{2\pi L_z}\sum_{\mathbf{G_z,G_z'}} W_{\mathbf{G_z,G_z'}}(q)\,,
\end{equation}
where $W_{\mathbf{G_z,G_z'}}(q)$ is the usual screened Coulomb interaction, computed using the Coulomb truncation technique, as suggested in Ref.\cite{PhysRevB.93.235435}.
The calculation of the screened Coulomb interation $W_{\mathbf{G_z,G_z'}}(q)$  was performed using the Abinit code for the untwisted, fully-relaxed, bilayer system. We used a dense 32x32 \textbf{k}-point grid, a supercell size $L_z=24$\AA, with a cutoff energy of 13 Ha for the screening, and using 500 bands for computing the dielectric function.
With this, we obtained an effective thickness $d=35$\AA which is a bit less than four times the atom-to-atom distance for the bilayer system, with the experimental value of $\epsilon=13.6$. This value was then used to evaluate the potential $V_{2D}(\mathbf{q}_{\parallel})$ in evaluating the effective $U$ and $V$ for the twisted bilayer system. 

In order to evaluate the Rytova-Keldysh potential with FFTs for the non-periodic Wannier orbitals, one resort to a Coulomb integral technique. This implies to compute the Fourier transform of $V_{2D}(s)$ on a sphere of finite radius $R$. The analytical solution for the vanishing in-plane momentum $\mathbf{G}_\parallel=0$ is given by $ V_{2D}^{\rm cutoff}(\mathbf{G}_\parallel=0) = \pi^2\rho_0\Big[ \frac{R}{\rho_0}\Big(H_1\Big(\frac{R}{\rho_0}\Big) - Y_1\Big(\frac{R}{\rho_0}\Big) \Big)-\frac{2}{\pi}\Big]\approx 2\pi R^2 \frac{[1-2\gamma-2{\rm log}(\frac{R}{2\rho_0})]}{4\rho_0}$, where $\gamma$ is the Euler-Mascheroni constant. The last expression is obtained using the fact that $R\ll \rho_0$ in the present work.
Using the above expressions, we obtained $U=1.448$meV and $V=0.145$meV, leading to a ratio $U/V \sim 9.99$. 

\af{
{\it FRG calculations} --- We address electronic order beyond DFT+U+V by employing the functional renormalization group (FRG)~\cite{metznerFunctionalRenormalizationGroup2012} that allows to resolve particle-particle and particle-hole instabilities in an unbiased manner.
We employ the static four-point truncated unity FRG (TUFRG) approximation in which the flow equations for the two-particle vertex function $\Gamma = \Gamma^{(4)}$ read
\begin{equation}
\partial_{\Lambda} \Gamma^{\Lambda} = \sum_\gamma \partial_{\Lambda} \gamma^{\Lambda} \,, \quad \partial_{\Lambda} \gamma^{\Lambda} =  \Gamma^{\Lambda} \circ \partial_{\Lambda} L^{\gamma, \Lambda} \circ \Gamma^{\Lambda} \,,
\label{eq-frg}
\end{equation}
where $\gamma \in \{P,D,C\}$ denotes the decomposition into two-particle reducible interaction channels that capture fluctuations accumulated in the particle-particle ($P$), direct particle-hole ($D$) and crossed particle-hole ($C$) channel. The onset of an ordered phase is signaled by the divergence of the two-particle vertex $\Gamma^{(4)}$ at a particular (critical) scale $\Lambda_\mathrm{c}$. The leading fermionic bilinear in the divergent channel is obtained by an eigenvalue decomposition of the channel-specific vertex $\gamma$
\begin{equation}
\gamma_{\kappa \kappa'}(\bvec q_\gamma) = \sum_i \phi^L_{\kappa,i}(\bvec q_\gamma) \, \gamma_i(\bvec q_\gamma) \, \phi^R_{\kappa',i}(\bvec q_\gamma) \,,
\label{eq-fermionic-bilinear}
\end{equation}
where $\kappa$ refers to the fermionic momentum variable $\bvec k_\gamma$. The left/right eigenvectors to the channels' eigenvalues $\gamma_i$ are denoted by $\phi^{L/R}_{\kappa,i}$ and give insight into the momentum structure of the order parameter. The critical scale $\Lambda_c$ serves as a proxy for the critical temperature of the transition. 

The FRG calculations are carried out with the
TUFRG backend of the divERGe code base~\cite{profe2024diverge}. The flow equations are solved using an adaptive Euler integrator and the form factor expansion of the (slow) fermionic momenta $(\bvec k_\cidx, \bvec k'_\cidx)$ is cut off at a distance of 3.01 $|\bvec a_M|$ in the parametrization of the vertex tensors. The bosonic momentum variable in each channel is resolved on a discrete mesh of $N_{\bvec k} = 60 \times 60$ momentum points in the Brillouin zone. An additional $N_{\bvec k_f}= 30 \times 30$ fine momentum mesh (for each coarse momentum point) is added to resolve the single-particle dispersion entering the (scaled) two-particle propagators $\partial_{\Lambda} L^{\gamma, \Lambda}$.
}
\nocite{*}
\bibliography{ref}

\end{document}